\documentclass[fleqn,usenatbib]{mnras}
\usepackage{newtxtext,newtxmath}
\usepackage[T1]{fontenc}

\DeclareRobustCommand{\VAN}[3]{#2}
\let\VANthebibliography\thebibliography
\def\thebibliography{\DeclareRobustCommand{\VAN}[3]{##3}\VANthebibliography}

\usepackage{graphicx}	    
\usepackage{amsmath}	    
\usepackage{upgreek}        
\usepackage{threeparttable}
\usepackage{tabulary}
\usepackage{caption}
\usepackage{orcidlink}

\newcommand{\DMunits}{\,pc\,cm$^{-3}$}
\newcommand{\RMunits}{\,rad\,m$^{-2}$}

\def\software#1{\texttt{#1}}

\newcommand{\Swin}{Centre for Astrophysics and Supercomputing, Swinburne University of Technology, P.O. Box 218, Hawthorn, VIC 3122, Australia}

\newcommand{\ATNF}{Australia Telescope National Facility, CSIRO, Space and Astronomy, PO Box 76, Epping, NSW 1710, Australia}

\newcommand{\ozgrav}{ARC Centre of Excellence for Gravitational Wave Discovery (OzGrav), Hawthorn, VIC 3122, Australia}

\newcommand{\NUIG}{National University of Ireland Galway, University Road, Galway, H91 TK33, Ireland}

\title[Repetitions of FRB\,20201124A]{Circularly polarized radio emission from the repeating fast radio burst source FRB\,20201124A}

\author[Kumar et al.]{P.~Kumar\orcidlink{0000-0003-1913-3092},$^{1}$\thanks{E-mail: pravirkumar@swin.edu.au}
R.~M.~Shannon\orcidlink{0000-0002-7285-6348},$^{1}$
M.~E.~Lower\orcidlink{0000-0001-9208-0009},$^{1,2}$
S. Bhandari\orcidlink{0000-0003-3460-506X},$^{2}$
A.~T.~Deller\orcidlink{0000-0001-9434-3837},$^{1}$
C.~Flynn\orcidlink{0000-0003-1110-0712}$^{1,3}$
\newauthor
and E.~F.~Keane\orcidlink{0000-0002-4553-655X}$^{4}$
\\
$^{1}$\Swin\\
$^{2}$\ATNF\\
$^{3}$\ozgrav\\
$^{4}$\NUIG
}

\date{Accepted 2022 March 7. Received 2022 March 7; in original form 2021 September 23}
\pubyear{2022}

\begin{document}
\label{firstpage}
\pagerange{\pageref{firstpage}--\pageref{lastpage}}
\maketitle

\begin{abstract}
The mechanism that produces fast radio burst (FRB) emission is poorly understood. Targeted monitoring of repeating FRB sources provides the opportunity to fully characterize the emission properties in a manner impossible with one-off bursts. Here, we report observations of the source of FRB\,20201124A, with the Australian Square Kilometre Array Pathfinder (ASKAP) and the ultra-wideband low (UWL) receiver at the Parkes 64-m radio telescope (\textit{Murriyang}). The source entered a period of emitting bright bursts during early 2021 April. We have detected 16 bursts from this source. One of the bursts detected with ASKAP is the brightest burst ever observed from a repeating FRB source with an inferred fluence of $640\pm70$ Jy~ms. Of the five bursts detected with the Parkes UWL, none display any emission in the range 1.1--4 GHz. All UWL bursts are highly polarized, with their Faraday rotation measures (RMs) showing apparent variations. We obtain an average RM of $-614$\,rad\,m$^{-2}$ for this FRB source with a standard deviation of $16$\,rad\,m$^{-2}$ in the UWL bursts. In one of the UWL bursts, we see evidence of significant circularly polarized emission with a fractional extent of $47\pm1$ per cent. Such a high degree of circular polarization has never been seen before in bursts from repeating FRB sources. We also see evidence for significant variation in the linear polarization position angle in the pulse profile of this UWL repeat burst. Models for repeat burst emission will need to account for the increasing diversity in the burst polarization properties.
\end{abstract}

\section{Introduction}\label{sec:intro}
Fast radio bursts (FRBs) are energetic $\upmu$s--ms-duration radio transients \citep{Cordes:2019, Petroff:2022} that originate at up to cosmological distances, with spectroscopic redshifts confirmed up to $z=0.66$ \citep{Chatterjee:2017, Bannister:2019_localization, Ravi:2019_localization}. The dispersion measures of these bursts lead to inferred redshifts of as high as $z=3$ in the currently known sample \citep{Bhandari:2018, CHIME:2021}. More than 600 distinct FRB sources are catalogued on the Transient Name Server (TNS\footnote{\url{https://www.wis-tns.org/}; visited 2021 September 1.}), but only a small proportion ($\approx$\,4 per cent) of these have been found to emit repeat bursts \citep{Spitler:2016, CHIME:2019, CHIME:2019_8repeaters, Kumar:2019, Fonseca:2020, Luo:2020, Kumar:2021, Bhardwaj:2021}. While several mechanisms have been proposed to explain the observed properties of these repeat bursts \citep{Platts:2019}, the true underlying physics behind these bursts remains an unsolved question.

Observing campaigns on repeating sources with interferometers enable localization and host-environment studies \citep{Chatterjee:2017}. However, out of 15 known host galaxies (HGs) of FRBs, only five of the associations are with repeating sources. A preliminary analysis of the two classes of source HGs indicated that the HGs of repeating FRB sources are less massive \citep{Bhandari:2020}. Furthermore, this small sample of localized repeating sources itself shows evidence for a diversity of progenitor environments \citep{Heintz:2020}. Studies comparing the hosts to determine whether progenitors of FRBs are consistent with a population of magnetars have resulted in contradiction \citep{Safarzadeh:2020, Bochenek:2021}. FRB\,20121102A originates from a highly star-forming dwarf galaxy with a coincident compact radio source consistent with a young magnetar progenitor scenario \citep{Chatterjee:2017, Metzger:2017}. Other repeating FRB HGs are less readily interpreted in this formalism -- the source of FRB\,20180916B is in a nearby spiral galaxy but offset from a star-forming knot \citep{Marcote:2020, Tendulkar:2021}, while the low-luminosity source FRB\,20200120E is associated with a globular cluster in the nearby galaxy M81 \citep{Bhardwaj:2021, Kirsten:2022}. There is also observational evidence against FRBs being produced by isolated magnetars. The discovery of modulation in the activity windows of repeating FRBs, such as the $16.35$-d periodicity observed in the case of FRB\,20180916B \citep{CHIME:2020_periodicity}, is also more readily interpreted in other progenitor models, such as high-mass X-ray binary systems \citep{Sridhar:2021, Tendulkar:2021}. Thus, a conclusive understanding of the progenitor location and its association with the HGs is not clear and cannot at this stage be used to argue for or against a given origin of the repeating FRBs.

There is emerging evidence for differences in the emission of repeating sources and the general FRB population. The repeating FRBs have long been suspected of having larger intrinsic pulse widths \citep{Keane:2016}. Using a well-controlled catalogue with a large sample of 535 bursts published by the Canadian Hydrogen Intensity Mapping Experiment (CHIME/FRB) with frequency coverage between 400--800 MHz \citep{CHIME:2021}, \citet{Pleunis:2021} showed that the bursts from repeating FRB sources have not only larger temporal widths but also narrower fractional bandwidths than those from apparent non-repeaters. However, it is unclear if these morphological differences correspond to a different class of progenitors, given there is no such observed distinction in the dispersion measure (DM) or scattering time distributions of these bursts \citep{Pleunis:2021}. Assuming the local environments of the burst sources are comparable, the lack of difference in the DM distribution suggests a similar luminosity function since the bursts are arising at the same distances.

FRBs have been observed to have a diversity of polarimetric properties and with tentative evidence for differences in the properties between the repeating and the apparently non-repeating bursts. Some of these bursts have linear polarization fractions up to 100 per cent \citep{Masui:2015, Michilli:2018, Gajjar:2018, Day:2020}. Some have significant circular polarization \citep{Petroff:2015, Caleb:2018, Cho:2020, Day:2020}. Others appear unpolarized, possibly due to propagation effects and limited frequency resolution of the observing instrument \citep[a higher rotation measure than can be detected;][]{Michilli:2018}. Some sources exhibit a flat linear polarization position angle (PA) across their burst profiles \citep{CHIME:2019_8repeaters, Fonseca:2020}. In contrast, others have shown PA swings across pulses, indicative of rotation, similar to what is seen in Galactic pulsars \citep{Cho:2020, Luo:2020}. Three repeating sources with reported polarimetry, FRBs~20121102A \citep{Gajjar:2018}, 20190711A \citep{Kumar:2021} and 20180916B \citep{Nimmo:2021} show no significant circular polarization and exhibit flat PAs across the burst profiles suggesting these differences as a key discriminant between the two populations. In contrast, another repeating source, FRB\,20180301A, revealed bursts with PA swings varying across the burst profile \citep{Luo:2020}. A key distinction emerging from these studies is the non-detection of circularly polarized radio emission in bursts from repeating sources \citep{Dai:2021}. 

FRB\,20201124A is an extraordinarily bright and active repeating source. This CHIME-discovered source \citep{CHIME:2021ATel14497} showed an episode of high activity in 2021 April--May, with more than 200 bursts reported in that month with a variety of radio telescopes \citep{Kumar:2021ATel14502, Kumar:2021ATel14508, Xu:2021ATel14518, Law:2021ATel14526, Herrmann:2021ATel14556, Kirsten:2021ATel14605, Farah:2021ATel14676, Hilmarsson:2021b, Marthi:2021, Lanman:2022}. Reported bursts have been detected at frequencies from $400$~MHz to $1.5$~GHz. Like many repeating FRB sources, the emission appears to be band-limited. For example, while two bursts were detected with the 25-m telescope at Onsala Space Observatory at 1.4 GHz, no emission was detected at 330 MHz with the 25-m Westerbork RT1 antenna observing simultaneously \citep{Kirsten:2021ATel14605}. Additionally, no prompt emission was detected at optical \citep{Zhirkov:2021ATel14532} and X-ray \citep{Campana:2021ATel14523, O'Connor:2021ATel14525} wavelengths during this highly active state.

Multiple independent efforts using radio interferometers \citep{Day:2021ATel14515, Day:2021ATel14592, Law:2021ATel14526, Marcote:2021ATel14603,Wharton:2021ATel14538}  pinpointed the burst source to a massive ($\sim 2 \times 10^{10} M_\odot$) and dusty star-forming galaxy SDSS J050803.48+260338.0 at $z=0.1$ \citep{Fong:2021, Ravi:2021}. Continuum radio emission coinciding with this position was detected at 650 MHz with the upgraded Giant Metrewave Radio Telescope \citep[uGMRT;][]{Wharton:2021ATel14529} and at 3 and 9 GHz with the Karl G. Jansky Very Large Array \citep[VLA;][]{Ricci:2021ATel14549}. However, the non-detection in high-resolution VLBI observations \citep{Marcote:2021ATel14603} implies the source is $0.4-6$~kpc in size \citep{Fong:2021}. Accordingly, this continuum radio emission is not powered by a compact pulsar wind nebula or an active galactic nucleus (AGN) core associated with the FRB progenitor as has been seen for the ``persistent radio source'' associated with FRB\,20121102A \citep{Chatterjee:2017}.

Here we report on a campaign monitoring the repeating source FRB\,20201124A using the Australian Square Kilometre Array Pathfinder (ASKAP) and the 64-m Parkes radio telescope (also known as \textit{Murriyang}) during its high-activity state. In Section~\ref{sec:observations}, we describe the follow-up observations and search methods used to find repeat bursts. In Section~\ref{sec:repeats}, we present the properties of the detected repeat bursts. We report the discovery of circular polarization from the source along with several repeat bursts. In Section~\ref{sec:discussion}, we discuss the inferred burst properties and implications on the emission mechanism of fast radio bursts.

\section{Observations and data processing}\label{sec:observations}
\begin{figure*}
\centering
\includegraphics[width=0.99\textwidth]{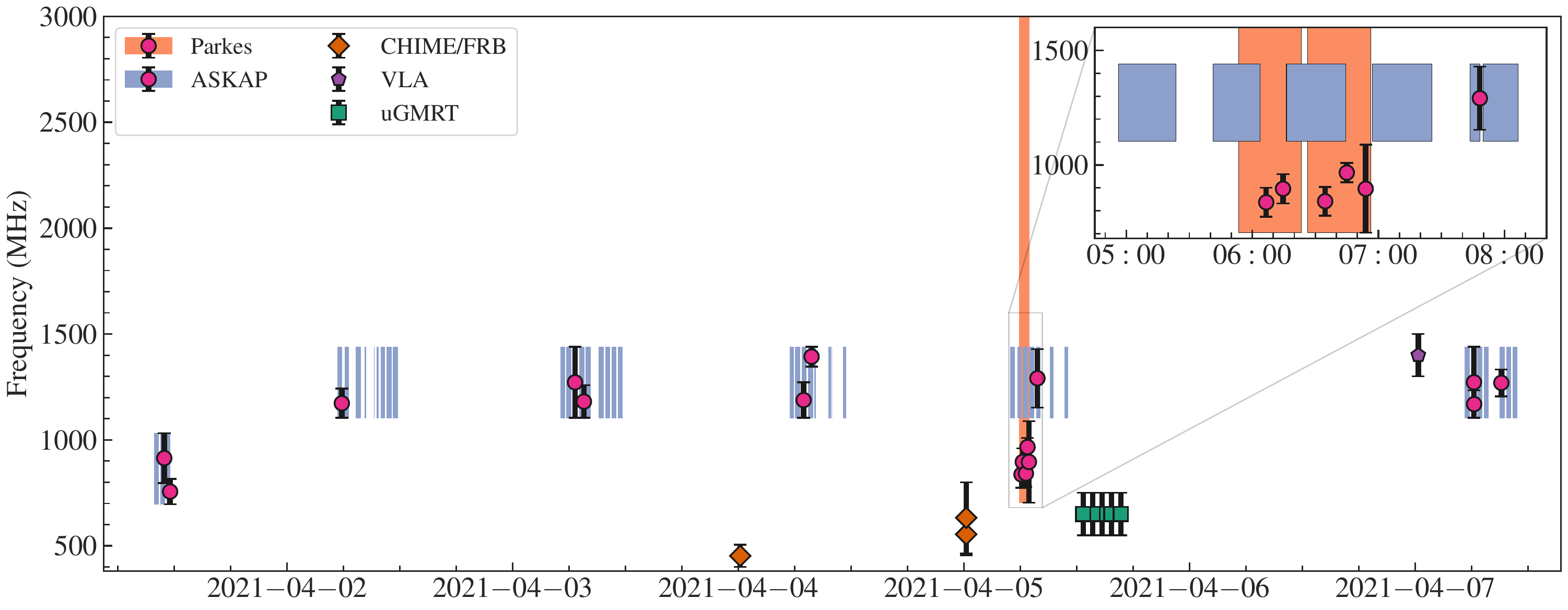}
\caption{Timeline of the follow-up observations for FRB\,20201124A. Also shown are bursts reported with CHIME \citep[three bursts;][]{Lanman:2022}, VLA \citep[one burst;][]{Law:2021ATel14526} and uGMRT \citep[48 bursts;][]{Marthi:2021} between April 1 and 8. Burst markers are overplotted at the zoom level and not resolvable in the case of uGMRT bursts and the two CHIME bursts on April 5. The simultaneous observation session with Parkes is magnified in the inset panel. The error bars represent the spectral envelope of the individual bursts.}
\label{fig:timeline}
\end{figure*}

On 2021 March 31, the CHIME/FRB collaboration reported both the discovery of the repeating FRB source 20201124A and that it had entered a period of increased activity. The bursts were detected toward the J2000 position R.A. = $05^{\rm{h}}08^{\rm{m}}$ and Decl. = $+26\degr11\arcmin$ at a dispersion measure of $413.5 \pm 0.5$ \DMunits \citep{CHIME:2021ATel14497}. On 2021 April 1, we began monitoring the source position with ASKAP. For the initial observations, we pointed the ASKAP antennas boresight at the above position. For later observations (April 4--7), we pointed one of the central ASKAP beams at an interferometrically derived position \citep{Day:2021ATel14515}, i.e. R.A. = $05^{\rm{h}}08^{\rm{m}}03.662^{\rm{s}}$ and Decl. = $+26\degr03\arcmin39.82\arcsec$ (J2000.0 epoch). We also observed the source with the 64-m Parkes radio telescope for $1$\,h on April 5, simultaneously with ASKAP. Figure~\ref{fig:timeline} shows a timeline of our follow-up observations and burst detections. Alongside, we also show FRB events reported from this source by the CHIME \citep{Lanman:2022}, the VLA \citep{Law:2021ATel14526} and the uGMRT \citep{Marthi:2021} in our observation time frame.

\subsection{ASKAP searches}\label{sec:askap}
Targeted monitoring of the FRB\,20201124A source with ASKAP was conducted as part of the Commensal Real-time ASKAP Fast Transients \citep[CRAFT\footnote{\url{https://astronomy.curtin.edu.au/research/craft/}};][]{Macquart:2010} project. Observations were carried out by incoherently summing the total intensities from each antenna for each of the 36 beams produced by the phased-array feed receivers. In this mode, the detection system is sensitive to a fluence of $3.7$~Jy~ms for a burst width of 1 ms and a threshold signal-to-noise ratio (S/N) of $10$ for an array observing with $36$ antennas; often smaller sub-arrays are used, and the sensitivity scales by $\sqrt{N_{\rm antennas}}$ for the incoherently summed array. The data were searched in near real time using the GPU-based detection system \software{fredda} \citep{fredda_ascl}. We identified 11 repeat bursts above an S/N of $9$ in a total of $16.5$~h of follow-up observations. The online voltage download system was triggered for five of these bursts, which enabled the capture of high-resolution baseband data. Details of the detection system and the trigger criteria used to download voltages can be found in the supplementary material of \citet{Bannister:2019_localization}. These baseband data were used to interferometrically localize the source \citep{Day:2021ATel14515, Day:2021ATel14592} to its host galaxy, SDSS J050803.48+260338.0 that has a redshift of $z = 0.098(2)$ \citep{Kilpatrick:2021ATel14516}. \citet{Fong:2021} reported a detailed study of the host galaxy and initial analysis of burst parameters for three of these ASKAP repeat bursts (A02, A03, and A07).

Here, we analyse the properties of all 11 repeat bursts using the available incoherently summed data. The data were saved to disc in \software{sigproc} filterbank format \citep{Lorimer:2011_ascl} with $336$ frequency channels at a time resolution of $1.2$~ms, and  $1$-MHz spectral resolution. Initial observations for $1.5$~h (on April 1) were conducted at a central frequency of $863.5$~MHz (low-band), whereas the rest of the observations were centred at 1271.5 MHz (mid-band). We recorded data with 22--24 antennas for all the ASKAP observations, except for the observations of burst A08 when only seven antennas were used.

\subsection{Parkes UWL searches}
We used the ultra-wideband low (UWL) receiver at Parkes, covering a continuous frequency range from $704$ to $4032$~MHz. The data were sampled with a time resolution of $64$~$\upmu$s in $6656$ frequency channels ($500$-kHz resolution) with each channel coherently dedispersed \citep{Hobbs:2020} for a DM of 412\,\DMunits. The data were stored in an eight-bit sampled \software{psrfits} search-mode file \citep{psrchive} with four polarization products. We performed single-pulse searches, by sub-banding the data into bandwidths of size 1$\times$3328\,MHz, 2$\times$1664\,MHz, 4$\times$832\,MHz, 8$\times$416\,MHz, 13$\times$256\,MHz, 26$\times$128\,MHz and 52$\times$64\,MHz. Additionally, we formed sub-bands by joining the bottom and top half portions of the adjacent sub-bands to be sensitive to overlapping signals. We searched the sub-banded data independently using the GPU based single-pulse search software \software{heimdall\footnote{\url{https://sourceforge.net/projects/heimdall-astro/}}} \citep{Barsdell:2012PhDT}. The search method and parameters is described in \citet{Kumar:2021}. We visually inspected all $\sim1500$ candidates in the DM range 350--500\DMunits~in a total of 13 sub-band searches and found five repeat bursts in the hour-duration observation above an S/N of $9$.  

A 2-min observation of a pulsed linearly polarized noise diode was performed for polarization calibration at the beginning of the observing session on April 5. We used earlier observations of radio source PKS~0407$-$658 for flux calibration. We corrected for polarimetric leakage using an instrumental response model provided by the observatory, based on observations of the millisecond pulsar PSR~J0437$-$4715 \citep{vanStraten:2004}.

\section{Results}\label{sec:repeats}
\begin{figure*}
\centering
\includegraphics[width=1.0\textwidth]{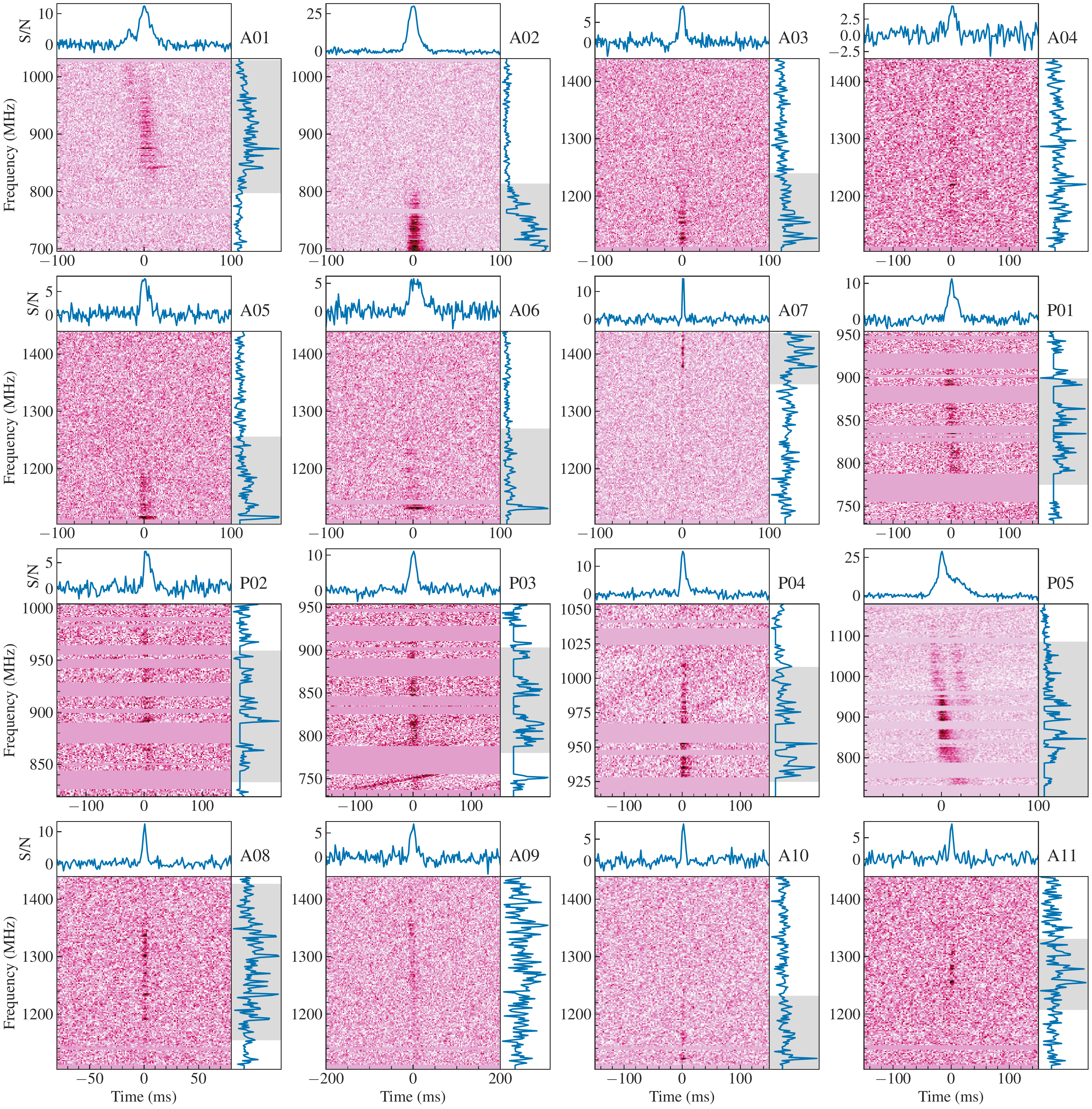}
\caption{
Dynamic spectra of repeat bursts detected with ASKAP and Parkes from FRB\,20201124A source. The bursts are plotted in chronological order.  Burst arrival times can be found in Table \ref{tab:burst_properties}. All bursts have been dedispersed to the best-fitting DM of 413 \DMunits. In each sub-plot, the bottom panel shows the dynamic spectrum (frequency resolution $\sim$ 1~MHz, time resolution $\sim$ 1~ms), and the right-hand side of the panel shows the time-averaged on-pulse spectrum. The grey band in these panels denotes the best-fitting sub-band in frequency. The top panel shows the frequency-averaged pulse profile using the best-fitting sub-band. The dynamic spectra are normalized, and intensity values are saturated at the fifth percentile. \label{fig:repeaterplots}}
\end{figure*}

In order to measure the burst properties, we extracted the data around the repeat bursts (total-intensity Stokes I for ASKAP bursts and all Stokes parameters for Parkes/UWL) in a \software{psrfits}-format file using the \software{dspsr} package \citep{dspsr}. Hereafter, we refer to the ASKAP detected bursts chronologically as A01--A11 and the Parkes bursts as P01--P05.  The dynamic spectra of all repeat bursts are presented in Figure \ref{fig:repeaterplots}. All of the UWL bursts are detected in the lower part ($\leq $ 1.1 GHz) of the band, which is predominantly contaminated by radio frequency interference (RFI). We mask channels whose statistical moments (variance, skewness, and kurtosis) deviate excessively from the median values across the whole band. This results in up to 50 per cent of the channels being flagged below 1.1 GHz. We take a more extended cut out of the data around the bursts to measure these average statistical moments.

\subsection{Dispersion Measure}
One of the fundamental properties of emission from the FRB sources is the dispersion, with its strength parametrized by the dispersion measure (DM), a measure of the electron column density along the line of sight. The difficulty in making an accurate determination of an FRB DM is compounded by the spectro-temporal evolution seen in the burst emission \citep{Platts:2021}, which can be degenerate with DM and lead to ambiguities~\citep{Hassall:2012}. Here we outline how we measured the DM of these repeat bursts.

\begin{table*}
\caption{FRB\,20201124A repeat burst parameters. Except for the detection S/N, all other burst properties are measured for the best-fitting sub-band after dedispersing to the DM $= 413$~\DMunits. The detection S/N is the value as reported by the search pipelines used for each instrument's data (see Section~\ref{sec:observations}).} 
\label{tab:burst_properties}
\small
\centering
\begin{threeparttable}
\begin{tabulary}{\textwidth}{LCCCCCCCCCCC}
\hline
Burst\tnote{(a)} & TNS event & TOA\tnote{(b)} & Detection & DM$_\mathrm{S/N}$\tnote{(d)} & DM$_\mathrm{Struct}$\tnote{(e)} & $\nu_\mathrm{low}$ & $\nu_\mathrm{high}$ & Width\tnote{(f)} & Gaussian & Fluence & S\tnote{(h)} \\
& FRB & (MJD) & S/N\tnote{(c)} & \DMunits & \DMunits & (MHz) & (MHz) & (ms) & S/N\tnote{(g)} & (Jy ms) & (Jy) \\
\hline
A01 & 20210401B & 59305.45386902 & 26.6 & 421 (2) & 413 (1) & 796  & -    &  18.1(7) & 34.2 & 125(14)  & 8(1) \\
A02* & 20210401A & 59305.47976873 & 60.1 & 413 (1) & 411.8 (6) & -  & 810  &  12.1(2) & 85.8 & 640(70) & 52(6) \\
A03* & 20210402A & 59306.24080906 & 13.4 & 415 (5) & - & -    & 1242 &  7.1(5)  & 19.8 & 51(6)  & 7(1) \\
A04 & 20210403A & 59307.27424089 & 10.7 & 423 (5) & - & -    & -    &  13(2)   & 8.8  & 21(4)  & 1.6(4) \\
A05 & 20210403B & 59307.31351659 & 14.3 & 417 (4) & - & -    & 1258 &  11(1)   & 18.9 & 58(7)  & 6(1) \\
A06 & 20210404C & 59308.28723624 & 12.3 & 420 (3) & - & -    & 1272 &  16(1)   & 15.4 & 51(6)  & 3(1) \\
A07* & 20210404B & 59308.32241382 & 12.2 & 411 (4) & - & 1348 & -    &  1.7(2)  & 17.5 & 27(3)  & 12(1)\\
A08* & 20210405C & 59309.32336795 & 16.7 & 414 (2) & - & 1153 & 1429 &  3.9(3)  & 18.8 & 41(5)  & 11(1)\\
A09 & 20210407A & 59311.25772833 & 14.4 & 424 (2) & - & -    & -    &  15(2)   & 14.5 & 35(5)  & 3(1)\\
A10 & 20210407B & 59311.25837971 & 9.1  & 413 (3) & - & -    & 1234 &  5.5(6)  & 12.1 & 29(4)  & 5(1)\\
A11* & 20210407C & 59311.37901451 & 9.6  & 409 (4) & - & 1206 & 1333 &  6.5(8)  & 12.6 & 28(4)  & 5(1)\\
P01 & 20210405D & 59309.25276249 & 26.4 & 418 (2) & 418 (2) & 774  & 900  &  18(1)   & 22.6 & 18(1)  & 1.2(1) \\
P02 & 20210405E & 59309.25831658 & 24.5 & 414 (3) & 415 (3) & 832  & 960  &  14(1)   & 15.3 & 13(1)  & 0.9(2) \\
P03 & 20210405F & 59309.27223804 & 26.6 & 415 (2) & 412 (1) & 779  & 904  &  12.7(7) & 23.3 & 16(1)  & 1.3(2) \\
P04 & 20210405G & 59309.27939324 & 31.9 & 411 (2) & 412 (1) & 924  & 1009 &  9.3(5)  & 28.9 & 15(1) & 1.5(2) \\
P05 & 20210405H & 59309.28563453 & 98.9 & 418 (2) & 418 (1) & 704  & 1088 &  22.6(6) & 72.4 & 39(1) & 2.4(1) \\

\hline
\end{tabulary}
\begin{tablenotes}[flushleft]
    \item[(a)] {For highlighted ASKAP bursts, we also recorded the high-resolution baseband data. The results will be presented separately.}
    \item[(b)] {Barycentric arrival time (TDB) at $\nu=\infty$}
    \item[(c)] {Reported S/N by the search pipeline, \software{fredda} for ASKAP bursts (using full 336 MHz band) and \software{heimdall} for UWL (for the best sub-band detection).}
    \item[(d)] {DM obtained by maximizing S/N using \software{pdmp}. DM uncertainties are 1-$\sigma$.}
    \item[(e)] {DM obtained by maximizing the coherent power across the bandwidth with \software{DM\_phase}.}
    \item[(f)] {Burst width (FWHM), obtained by fitting the pulse profile with a Gaussian function.} 
    \item[(g)] {Pulse profile is convolved with a set of normalised Gaussian filters. Then the best template is used to calculate the integrated S/N.}
    \item[(h)] {Peak flux density. Uncertainties are dominated by beam corrections.}
\end{tablenotes}
\end{threeparttable}
\end{table*}

We use the \software{PSRCHIVE} tool \software{pdmp} to determine the DM that maximizes the S/N of all the repeat bursts. \software{pdmp} computes the S/N using boxcar matched filtering of the pulse profile. We identify by eye the frequency sub-band where the bursts are bright and use only those channels for DM determination. We also take a broader sub-band to include all the burst signals and confirm the obtained results. We find that the S/N-maximized DM covers a wide range from 409 to 423\DMunits~in our sample. These large deviations can be attributed to the low S/N, narrow fractional bandwidth, and the presence of sub-structure in some of the bursts. We also see evidence for sub-components present in the dynamic spectrum of A01 and P05 (see Figure \ref{fig:repeaterplots}). To consider these and align the burst structure properly in time, we use the \software{DM\_phase} package \citep{Seymour:2019}. This method uses the coherent power across the dynamic spectrum to maximize the burst structure instead of S/N. To estimate the DM uncertainties for both methods, we follow the approach prescribed in \software{DM\_phase} \citep{Seymour:2019}. We fitted the obtained S/N-DM curve \citep{Cordes:2003} with a higher degree (10) polynomial and used the peak to establish the DM for each burst. The uncertainty in DM is calculated by propagating the uncertainty given by the residuals of the polynomial fitting \citep{Platts:2021}. We report the measured DM with uncertainties for all repeat bursts in Table \ref{tab:burst_properties}. The low S/N and time resolution of ASKAP bursts A03--A11 resulted in poor fitting when optimizing the burst structure with the \software{DM\_phase}. For these bursts, we do not provide DM$_\mathrm{Struct}$ in Table \ref{tab:burst_properties}.

\begin{figure*}
\begin{center}
  \includegraphics[width=0.99\textwidth]{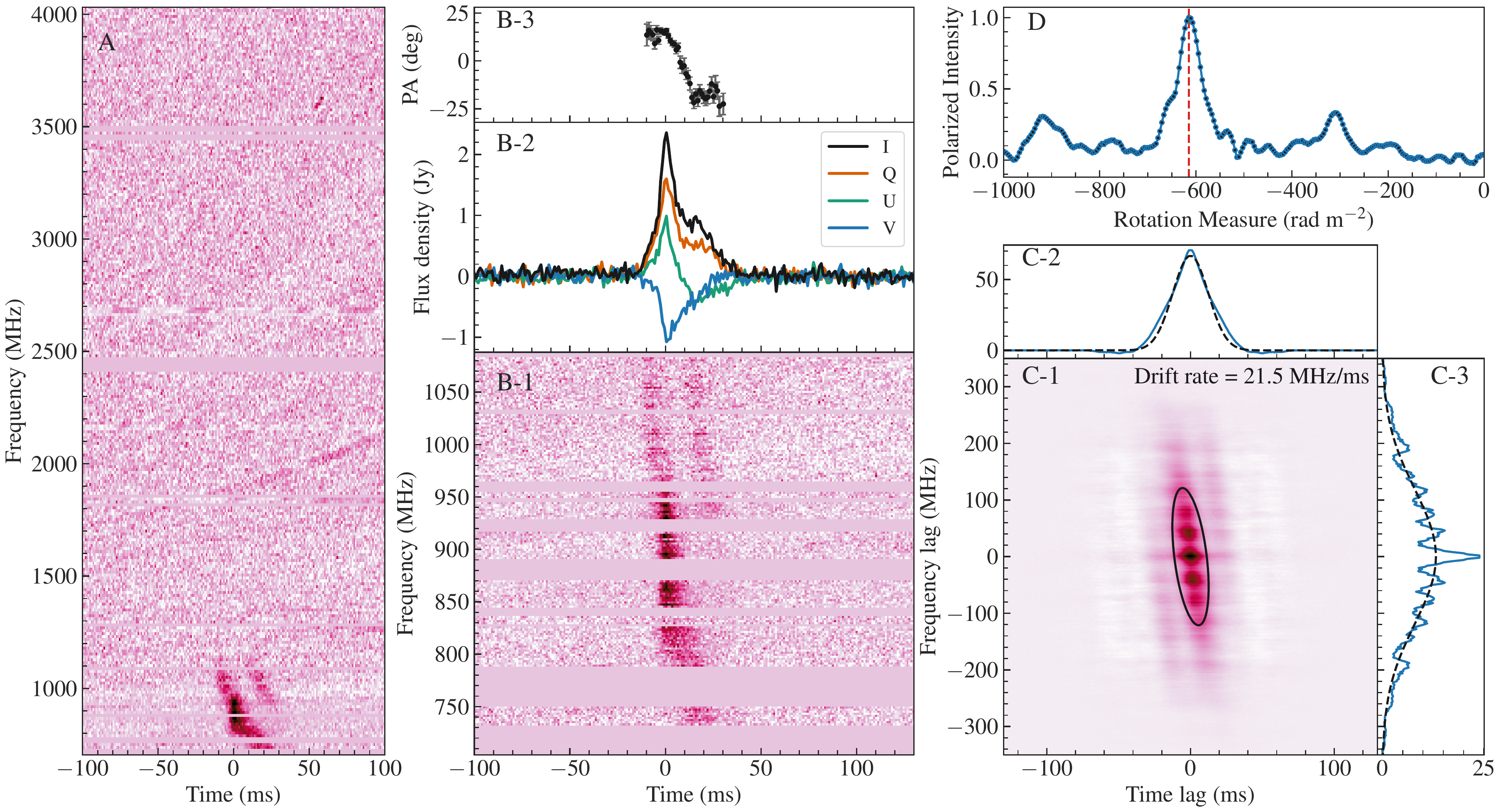}
\end{center}
\caption{Dynamic spectra and polarization profiles of the Parkes/UWL repeat burst P05 from FRB\,20201124A source. The data have been dedispersed to the best-fitting global DM of 413\DMunits. The dynamic spectra are normalized, and intensity values are saturated at the fifth percentile. {\em Panel A}: The dynamic spectrum (frequency resolution = 26~MHz, time resolution = 1~ms) across the full UWL band. {\em Panel B}: Panel B-1 shows the dynamic spectrum over the best sub-band (frequency resolution = 2~MHz). Panel B-3 and B-2 show, respectively, the polarization PA and frequency-averaged time series for the four Stokes parameters.  {\em Panel C}: Panel C-1 shows the autocorrelation function for the burst dynamic spectrum fitted with a 2D Gaussian. The zero-lag noise spike has been removed. Panels C-2 and C-3 display the average of the computed ACF (and the fitted Gaussian model) along the time and frequency axes, respectively. {\em Panel D}: The linear polarization intensity as a function of rotation measure computed with \software{RMFIT}. The best RM is at $-$614 \RMunits. 
\label{fig:p05_uwl}  }
\end{figure*}

We define a global DM for this FRB source as the weighted mean of DM$_\mathrm{Struct}$ measured for repeat bursts in our sample. Doing this, we obtain the global DM to be $413 \pm 1$ \DMunits. Our determined value of DM is in agreement and within 1-$\sigma$ of the value measured for a sample of bright and multi-component bursts detected using the 100-m Effelsberg radio telescope \citep{Hilmarsson:2021b} and within 2-$\sigma$ for the brightest burst detected using the uGMRT \citep{Marthi:2021}. We use the determined global DM for this source to measure the burst properties reported in Table \ref{tab:burst_properties} and throughout this paper. We find that the optimized DM of the brightest UWL burst P05 significantly deviates from the global DM (see Figure~\ref{fig:p05_uwl}), so in Section~\ref{subsec:P05_dm}, we analyse the P05 sub-structure considering different DM scenarios.

\subsection{Burst properties}
All five UWL and at least 9 out of the 11 ASKAP bursts are observed to be band-limited with emission bandwidths $\lesssim 350$~MHz. To characterize their properties, we extract the portion of the dynamic spectrum where the signal is present. In most cases, we estimate the spectral envelope by fitting the on-pulse average spectrum with a Gaussian function and extracting a band twice the measured full width at half-maximum (FWHM) around the peak signal. However, in other cases where the signal spectrum is flat, weak or contaminated with RFI, the baseline variations prohibit fitting a model, so we estimate the envelope by eye such that all signals are contained. The spectral envelope ($\nu_\mathrm{low}$ and $\nu_\mathrm{high}$) we used to measure the properties is in Table~\ref{tab:burst_properties} and also shown as a shaded region in Figure \ref{fig:repeaterplots}. In some of the ASKAP bursts, the signal either is across the whole band or extends beyond the edges; the lower or upper extent of the spectral envelope cannot be determined. For these bursts, the value of $\nu_\mathrm{low}$ or $\nu_\mathrm{high}$ is not given in Table~\ref{tab:burst_properties}.

To measure the temporal width, we use a least-square routine, fitting the frequency-averaged pulse profile with a single Gaussian function. A Gaussian model is sufficient for the pulse profile of all repeat bursts reported here.  We do not attempt to measure any associated scattering time-scale. Aside from bursts A01 and P05, we do not find any temporal sub-structure or evidence of drifting in the dynamic spectra. The low S/N and time resolution of ASKAP bursts restrict our ability to identify sub-structures in the burst. To measure the integrated S/N, we convolve the pulse profile with a set of square-normalised Gaussian templates for a range of widths using the package \software{spyden\footnote{\url{https://bitbucket.org/vmorello/spyden}}} and report the best-fitting template S/N. To determine fluence, we use twice the measured FWHM duration of the burst as the on-pulse region to include all of the burst power. For ASKAP detections, we obtain the flux density values using the radiometer equation for single pulses, assuming a system equivalent flux density of $1800$~Jy for each antenna as described in \citet{Bannister:2017}.

Initial ASKAP observations were pointed with antenna boresight at the less accurate position reported in \citet{CHIME:2021ATel14497}. Thus a correction to the flux density measured with the best-detection beam is needed. We assume a Gaussian beam model and follow the approach in \citet{Bannister:2017} to calculate the signal response at the FRB\,20201124A source position within the beam. We use the obtained response as a correction factor for the measured beam flux density. As in \citet{Bannister:2017} and \citet{Shannon:2018}, we conservatively assume a 10 per cent root-mean-square (RMS) variation for the beam gain and width to measure the uncertainties associated with the flux density correction. We also assume an RMS uncertainty of 1 arcmin in the beam centre positions. The beam-corrected fluence and peak flux density values are listed in Table~\ref{tab:burst_properties}. We pointed the Parkes using the more accurate interferometrically determined position, so we do not need to correct for any attenuation. We determine the uncertainties on flux density for the Parkes bursts using the RMS noise in the pulse profile.

\subsection{Polarization analysis}
The UWL bursts were polarization calibrated using the procedures detailed in \citet{Lower:2020}. Frequency channels known to be strongly affected by RFI and those affected by aliasing at the edges of the UWL $128$~MHz sub-bands were excised in the calibrator data \citep{Hobbs:2020}. To compensate for this, we interpolate and smooth the flux and polarization calibrator solutions using the tool \software{smint} from \software{psrchive} \citep{psrchive}.

\begin{table}
\caption{Rotation measure of UWL bursts obtained with \software{rmfit} and direct Stokes $Q$--$U$ fits (RM$_\mathrm{nest}$). Polarization fractions are obtained after correcting for individual RM$_\mathrm{nest}$. Uncertainties are given as 1-$\sigma$ confidence interval.} 
\label{tab:rm_table}
\small
\centering
\begin{tabulary}{0.99\columnwidth}{LCCCCCC}
\hline
Burst & $\Delta$T & RM$_\mathrm{fit}$ & RM$_\mathrm{nest}$ & P/I & L/I & V/I \\
      & (min)     &(\RMunits) & (\RMunits) &  & & \\
\hline
P01 & 0    & $-$636(7)  & $-$644(4) & 0.86(5) & 0.82(5) & $-$0.02(4)\\
P02 & 8.0  & $-$611(11) & $-$594(11)& 0.55(5) & 0.54(5) & 0.05(5)\\
P03 & 28.0 & $-$614(8)  & $-$616(3) & 0.82(5) & 0.78(5) & $-$0.20(4)\\
P04 & 38.3 & $-$607(4)  & $-$613(8) & 0.88(5) & 0.88(5) & $-$0.09(4)\\
P05 & 47.3 & $-$614(6)  & $-$612(1) & 0.89(2) & 0.76(2) & $-$0.45(2)\\
\hline
\end{tabulary}
\end{table}

\begin{figure*}
	\centering
\includegraphics[width=0.99\textwidth]{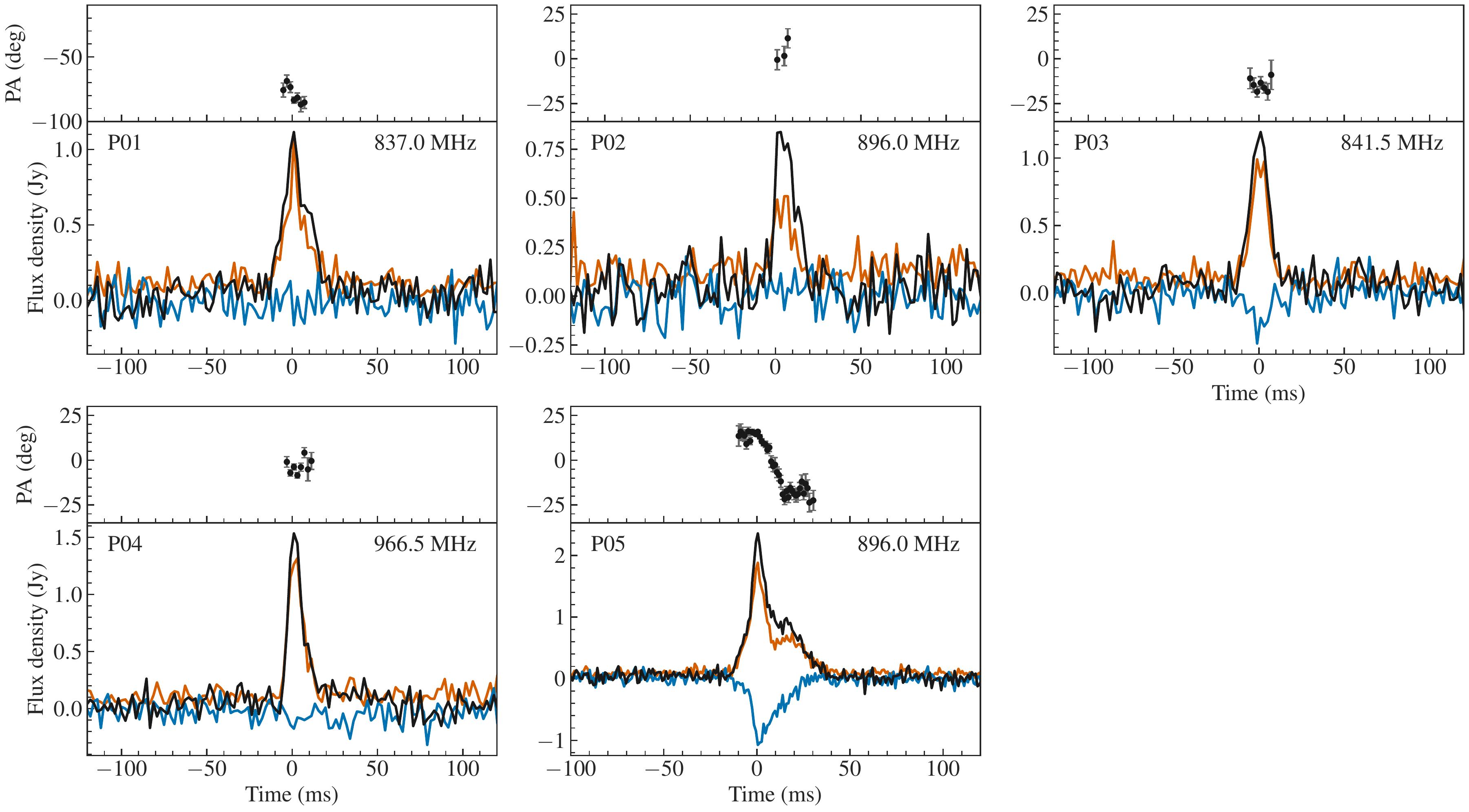} 
\caption{Polarization profile of the Parkes/UWL repeat bursts from the FRB\,20201124A source. {\em Top panel}: De-biased linear polarization PA versus time. {\em Bottom panel}: Frequency-averaged time series for the Stokes I (black), V (blue), and total linear polarization, L (orange). The data are corrected for the mean best-fitting RM of $-614$ \RMunits~using the centre of the best-fitting sub-band as the reference frequency (labelled in top right).}
\label{fig:polprofile}
\end{figure*}

We used two methods to search for Faraday rotation and ascertain its strength in the data. We first used the \software{rmfit} programme from \software{psrchive} to perform a brute-force search for a range of rotation measure (RM) values, $\pm 10^4$ \RMunits\ with a step size of 1 \RMunits.
We then used the Quadratic fitting algorithm \citep{Noutsos:2008} implemented in \software{rmfit} to obtain the best-fitting RM on a zoomed-in grid around the peak value of the brute-force search. Figure \ref{fig:p05_uwl} shows the polarization intensity as a function of rotation measure obtained from the \software{rmfit} brute-force search for the burst P05. Additionally, we used \software{rmnest\footnote{\url{https://github.com/mlower/rmnest}}} \citep{Lower:2020}, which utilizes Bayesian parameter estimation to obtain the RM for the UWL bursts by directly fitting the Stokes $Q$ and $U$ spectra of the calibrated data \citep{Bannister:2019_localization}. The resulting RM measurements with 1-$\sigma$ uncertainties using both methods are presented in Table \ref{tab:rm_table}. For further analysis, we now consider the best-fitting RM ($Q$--$U$) value of each burst and correct the UWL data for Faraday rotation.

We observe a large discrepancy in the measured RM values for the first two Parkes/UWL bursts (P01--P02), which are only 8 min apart. These bursts are heavily contaminated with RFI, and the paucity of unflagged frequency channels might have introduced additional systematic errors in the RM measurement. On the other hand, we find the RM of the other three bursts (P03--P05) consistent within 1-$\sigma$ uncertainties. The standard deviation of the burst RMs is $16$~\RMunits. We obtain a global source RM of $-614$~\RMunits~by taking the weighted mean of the best-fitting RM (Q--U) of all UWL bursts (P01--P05). We use this mean value as the FRB source RM throughout the paper unless otherwise mentioned. Assuming the RM values for P01--P02 correctly reflect the amount of Faraday rotation, we see a fractional variation of $\sim$3 per cent around the obtained global RM in the UWL bursts, all of which are detected merely within an hour. \citet{Hilmarsson:2021b} measured a fractional variation of $\sim$2 per cent in the RM around a mean value of $-601$\RMunits~in the Effelsberg repeat bursts detected on 2021 April 9. Thus, we find the RM determined for UWL bursts in agreement, given a similar significant RM variation in bursts from this FRB source around our observing span.

The Faraday-corrected spectra were averaged over frequency to obtain the polarimetric pulse profile. We determined the absolute linear polarization PA using the frequency-averaged Stokes $Q$ and $U$ for each burst. We use the methods described in \citet{Everett:2001} and \citet{Day:2020} to remove the bias in the total linear polarization, $L$. We use a 3-$\sigma$ threshold on the de-biased $L$ to establish significant measurements of PA. The frequency-averaged Stokes $I$, $L$, and $V$ profile of the UWL bursts along with the significant measurements of PA are plotted in Figure \ref{fig:polprofile}. We also measure the total $P/I$, the linear $L/I$, and the circular polarization fractions $V/I$ for each burst, where $ P = \sqrt{L^2 + V^2}$ is the total polarization. These are listed in Table 2. All five UWL bursts are highly polarized with a total polarization fraction of 60 per cent or greater. Circular polarization fractions up to 20 per cent have been reported from this source for the first time in a repeating FRB source \citep{Hilmarsson:2021b}. We also measure a significant amount (> 15 per cent) of fractional circular polarization in UWL bursts P03 and P05. However, we do not see any evidence of circular polarization in the interim burst P04 or in the first two UWL bursts (P01--P02).

\subsection{Drift-rate analysis}
Two bursts with high S/N (A01 and P05) show sub-structure drift in frequency when de-dispersed to the mean global DM obtained from the sample. This characteristic property of drifting (the ``sad trombone'' effect) has been observed earlier in bursts from other repeating FRB sources \citep{CHIME:2019_8repeaters, Hessels:2019}. We computed the 2D autocorrelation function (ACF) for the on-pulse dynamic spectrum of both bursts.  We removed the zero-lag noise spike in both time and frequency. We measured the linear drift rate, defined as the tilt in the ACF ellipse \citep{Hessels:2019} by fitting a generalized 2D Gaussian to the ACF and obtained 9 and 21.5\,MHz\,ms$^{-1}$ for bursts A01 and P05, respectively. We note that not all of the bursts in our sample show a drifting structure. This is consistent with the morphology of bursts detected from other repeating sources \citep{Pleunis:2021}. A spread of drift rates like the one determined here is not uncommon and is consistent with other measured drift rates for the bursts from this FRB source \citep{Marthi:2021}. The low-resolution of the available ASKAP data and the insufficient S/N of bursts curtail the possibility of further resolving the burst structure components.

\begin{figure}
\centering
\includegraphics[width=1\columnwidth]{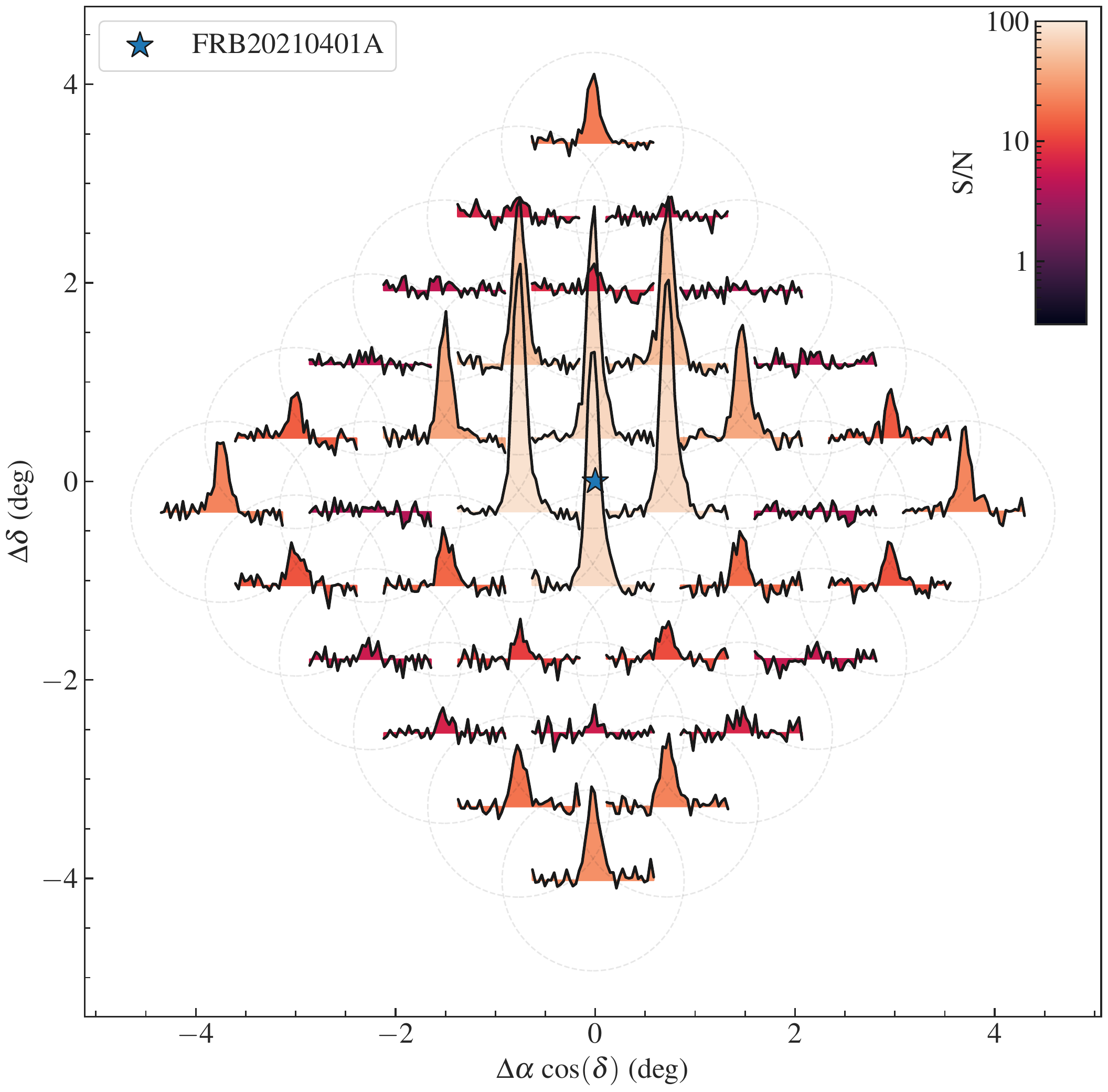} 
\caption{Pulse profiles of the ASKAP burst FRB20210401A (A02) for all 36 beams. The FRB detections across the beams map out the non-trivial point spread function of the ASKAP beam-former. The actual position of the FRB source is marked in blue. We show FWHM power (1.8\degr at 863.5 MHz) as dashed circles. The beam offsets in Right Ascension  and Declination, ${\rm{\Delta }}\alpha \cos \delta $ and ${\rm{\Delta }}\delta $, respectively, are measured relative to the burst interferometric position (see Section~\ref{sec:observations}).}
\label{fig:A02}
\end{figure}

\subsection{An extremely bright burst}
The second ASKAP-detected burst, A02, is the brightest repeat burst ever detected from an FRB source. The burst was sufficiently bright that it was detected with high significance (S/N > 10) in 20 of the total 36 ASKAP beams with strong detections in the side-lobes of the antenna response, as shown in Figure \ref{fig:A02}. We obtain almost comparable S/N $\sim$84 in two of the centred beams (0.8 deg offset from the FRB position), whereas the S/N in the beam closest to the burst (0.4-deg offset) is 68. The side-lobe beam detections (4 deg offset) were only a factor of 5 lower in S/N than the central beam detections. The similarity of the S/N in the central beam detections and the strong side-lobe detections (relative to the central beams) suggests that the primary beam detection had saturated a component of the ASKAP detection system, and the fluence calculated from the central beams was underestimated. Considering the possibility of saturation, we estimate the fluence from the neighbouring beam 5 (1.4-deg offset from the FRB position). We use ASKAP low-band holographic observations \citep{Hotan:2021} at the burst central frequency of 753 MHz to estimate the beam shape and the attenuation at the burst position. We measure the uncorrected fluence and attenuation factor for beam 5 to be $180\pm4$ Jy~ms and $0.28\pm0.03$ respectively; thus, we estimate the true burst fluence to be $640\pm70$ Jy~ms.

\begin{table}
\caption{FRB\,20201124A follow-up observations.} 
\label{tab:followupobs}
\small
\centering
\begin{threeparttable}
\begin{tabulary}{1\columnwidth}{LCCCC}
\hline
Instrument & Centre frequency & Sensitivity\tnote{{b}} & Obs. & Bursts\\
(reference)\tnote{{a}} & (MHz) & (Jy ms) & (h)\\
\hline 
ASKAP Low  & 863.5  & 4.53 & 1.2   & 2\\
ASKAP Mid  & 1271.5 & 4.53 & 15.3  & 9\\
UWL Low    & 1024   & 0.34 & 1     & 5\\
UWL Mid    & 1856   & 0.27 & 1     & 0\\
UWL High   & 3200   & 0.21 & 1     & 0\\
VLA Low (1, 2)    & 1400   & 0.5  & 1.73  & 1\\
VLA High (1, 2)   & 6000   & 0.13 & 6.93  & 0\\
Effelsberg (3)    & 1360   & 0.14 & 4     & 20\\
uGMRT (4)         & 650    & 1    & 3     & 48\\
DSN (5)           & 2260   & 5.4  & 2.97  & 0\\
Effelsberg (6)    & 6000   & 0.07 & 6     & 0\\
Stockert (7)      & 1380   & 25   & 90    & 1\\
\hline
\end{tabulary}
\begin{tablenotes}[flushleft]
    \item[{a}]  {(1) \citet{Ravi:2021}; (2) \citet{Law:2018}; (3) \citet{Hilmarsson:2021b}; (4) \citet{Marthi:2021}; (5) \citet{Pearlman:2021ATel14519}; (6) \citet{Spitler:2021ATel14537}; (7) \citet{Herrmann:2021ATel14556}.}
    \item[{b}] The limiting fluence for a pulse width of 1~ms and S/N threshold of 10.
\end{tablenotes}
\end{threeparttable}
\end{table}

\subsection{Burst rates}
We can combine our observations of the FRB\,20201124A source with those reported by others \citep{Xu:2021ATel14518, Law:2021ATel14526, Wharton:2021ATel14538} around our observation span (2021 April 1--7) to constrain the chromatic burst activity rate. More than 200 bursts from this source have been reported during 2021 April \citep{Hilmarsson:2021b, Marthi:2021, Piro:2021, Xu:2021ATel14518}, so we assume that the source was active across the whole span and our observations are entirely within a single active window. Several prolific repeating FRBs show strong clustering in their emission activity. Nevertheless, it has been well established for one of the most extensively studied sources, FRB\,20121102A, that the burst rate distribution roughly follows Poissonian statistics during its active phase \citep{Cruces:2021, Zhang:2021} after excluding events with separation $\lesssim$ 1 min. Since this is the case here in our observations, we assume a Poissonian rate distribution for the rate analysis.

Given the wide bandwidth of the Parkes/UWL, we divide it into a composite of three independent telescope sub-bands, similar to the RF bands described in \citet{Hobbs:2020} at a central frequency of 1.024, 1.856, and 3.2 GHz with a bandwidth of 640, 1024, and 1664 MHz, respectively. We also include all reported detections of repeat bursts and exposure times on the FRB\,20201124A source. The sensitivity limits and on-source follow-up time for each instrument with references are in Table \ref{tab:followupobs}.

\begin{figure}
	\centering
    \includegraphics[width=1\columnwidth]{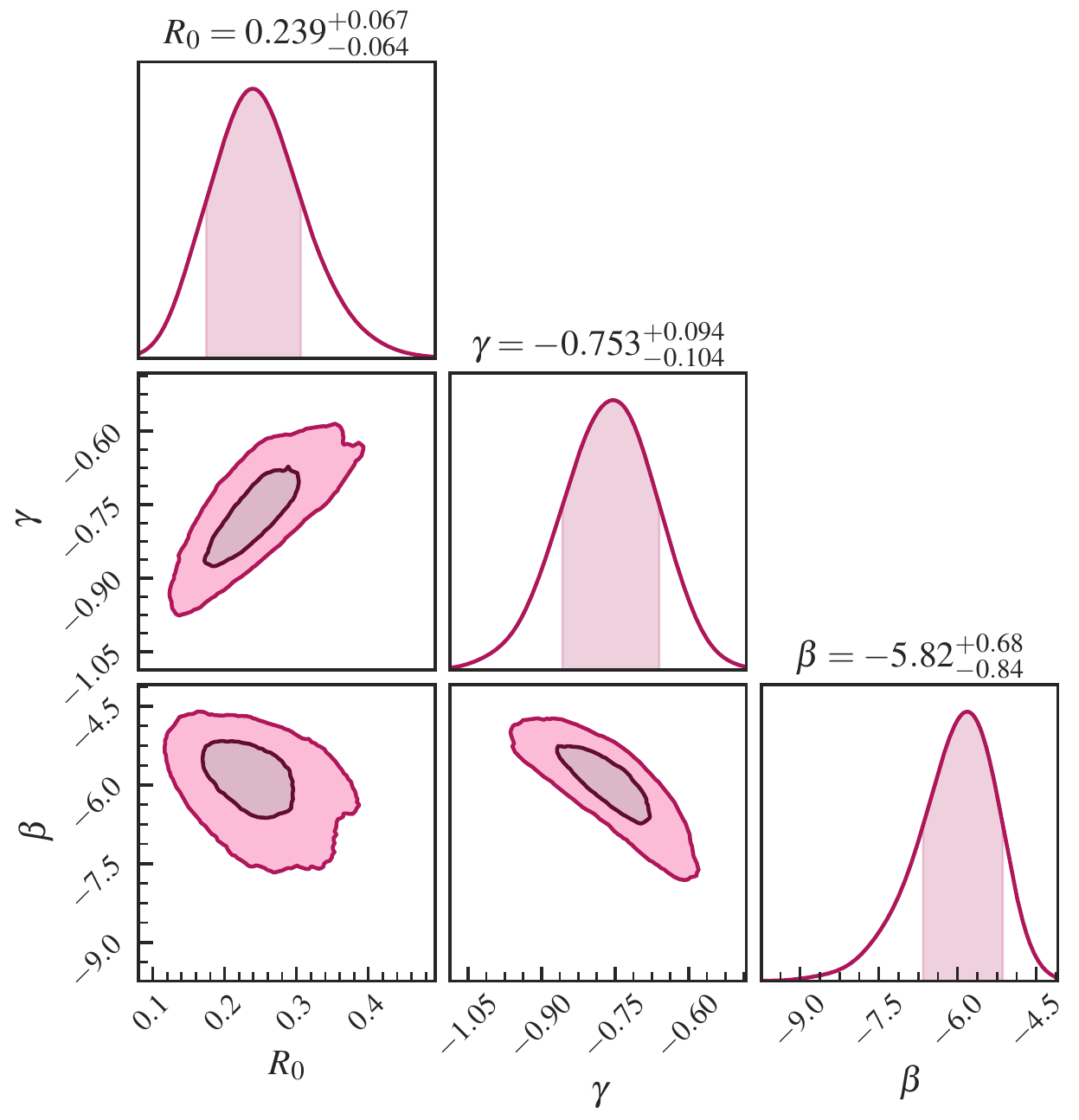} 
    \caption{Posterior distributions for burst rate parameters using all reported follow-up observations for the repeating source FRB\,20201124A. Top panels show the one-dimensional marginalized distributions for $R_0$, $\gamma$, and $\beta$, with 1-$\sigma$ confidence level as shaded region. $R_0$ has been scaled to ASKAP sensitivities ($S_0 = 4.5$ Jy~ms) and frequencies ($\nu_0 =$ 1.3~GHz). Other panels show the two-dimensional distributions of these parameters with the contours representing 1- and 2-$\sigma$ confidence levels for 2D Gaussians.}
    \label{fig:rates_post}
\end{figure}

We follow the formalism used in \citet{Kumar:2019} to estimate the cumulative burst rate for a survey using the parameters $R_0$ (rate of bursts above fluence $S_0$), $\beta$ (power-law spectral index) and $\gamma$ (cumulative power-law index for fluence). Using a Poissonian likelihood for the rate, we sample the posterior distribution with the nested sampling algorithm \software{multinest} \citep{Feroz:2009} implemented in \software{bilby} \citep{Ashton:2019}. We assume uniform priors on $\beta$ and $\gamma$ ($-10 < \beta, \gamma < 10$), and log-uniform priors on $R_0$ between $10^{-6}$ and $100$\,h$^{-1}$, where we use ASKAP mid-band search parameters for reference (frequency $\nu_0=1.3$~GHz and sensitivity $S_0=4.5$\,Jy\,ms). The obtained posterior distributions are shown in Figure \ref{fig:rates_post}. We thus constrain the burst rate of FRB\,20201124A to 0.24 h$^{-1}$ at 1.3 GHz for a fluence limit of 4.5~Jy~ms during its activity window. We find the spectral dependence of the burst rate to be very steep, $\beta \sim -6$. A separate analysis using a power law with an exponential cut-off in frequency for fluence did not return well-constrained results.

\subsection{Burst P05 sub-structure}\label{subsec:P05_dm}
The dynamic burst structures of FRBs depend on the accurate determination of DM and could easily result in over-interpretation \citep{Platts:2021}. Our earlier analysis of the Parkes/UWL repeat burst P05 is based on DM = 413\DMunits~that we obtain from the statistical average of a sample of bright bursts. At this DM, we can identify an upward frequency drift in the burst components in contrast with the downward drift observed in bursts from other repeating sources. However, when we take a closer look at this individual burst structure, we find that both the S/N and structure-optimized DM is 418\DMunits, which significantly deviates from the sample average. The burst intensity structure is clearly more complicated, so we attempt to understand it in this section.

\begin{figure}
    \centering
    \includegraphics[width=1.0\columnwidth]{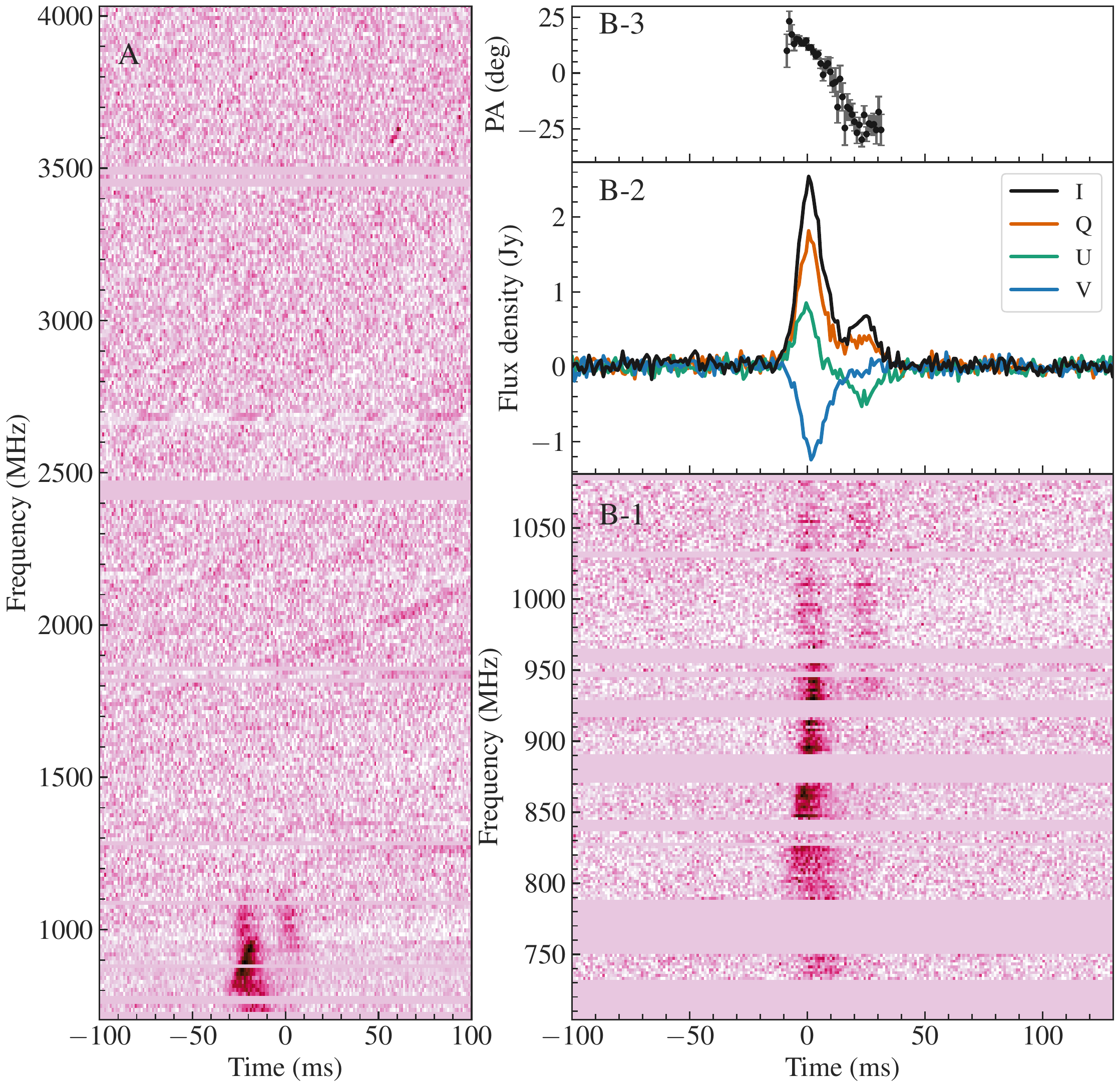}
    \caption{Intensity dynamic spectra and polarization profile of the Parkes/UWL repeat burst P05. Same as Figure \ref{fig:p05_uwl}, but now dedispersed to the DM of 418\DMunits.}
    \label{fig:P05_alternate}
\end{figure}

\begin{figure}
    \centering
    \includegraphics[width=1.0\columnwidth]{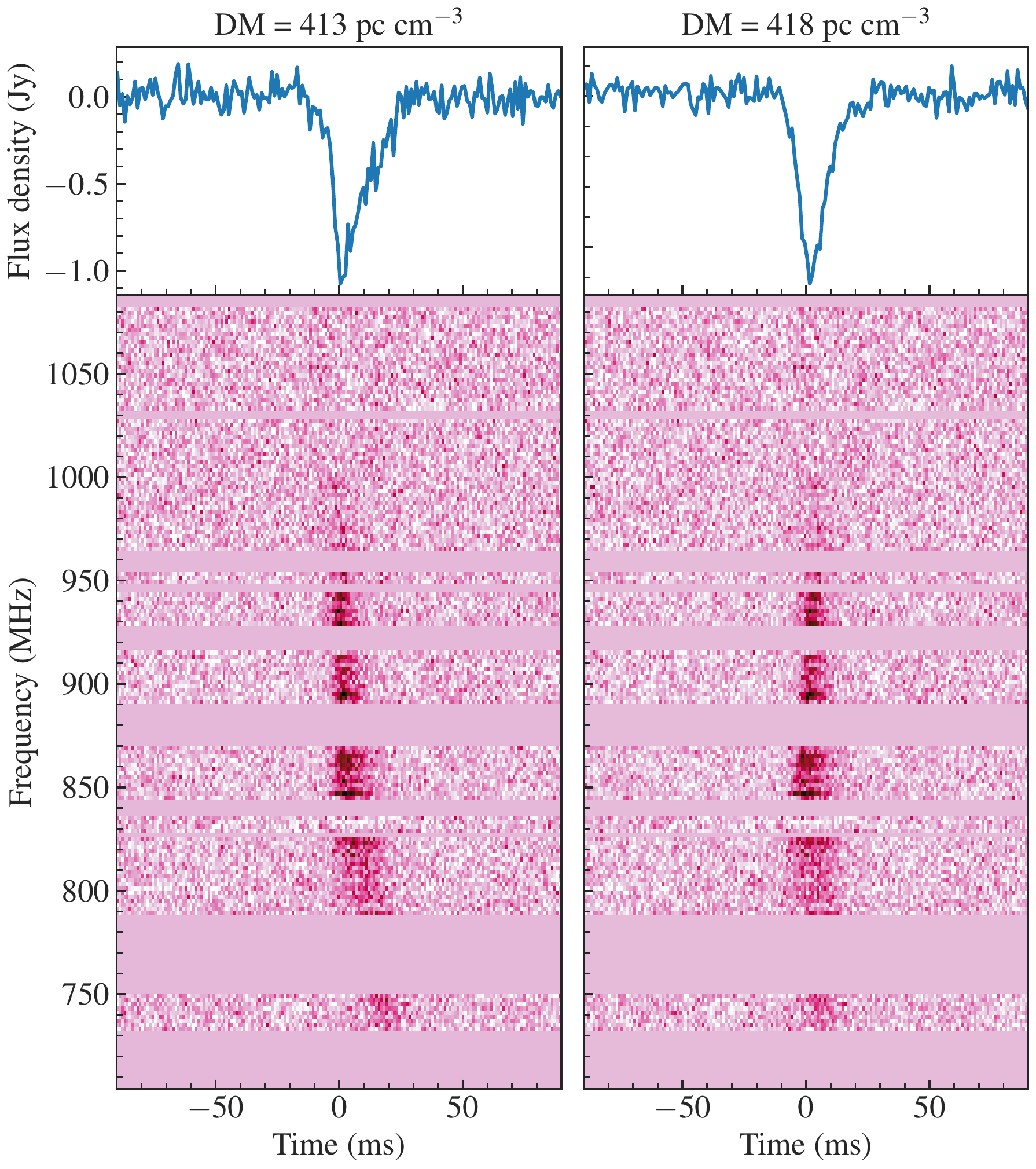}
    \caption{Stokes V dynamic spectra of UWL repeat burst P05 dedispersed at two extreme DMs. The spectra are plotted with 2 MHz spectral resolution and 1 ms temporal resolution. Normalized spectra values are saturated at the 95th percentile. The top panel shows the frequency-averaged profiles.}
    \label{fig:stokesvplot}
\end{figure}

We first use the individual burst-optimized DM = 418\DMunits~to measure properties for the UWL burst P05. The second upward drifting component is now separated in time and can be regarded as an entirely independent sub-pulse, possibly a post-cursor event. We use a two-component Gaussian model and obtain a peak separation of $21.2\pm0.5$ ms between the two sub-pulses (C1 and C2). The signal in the sub-pulse C2 is weak and has a comparatively smaller spectral envelope. We constrain the spectral emission in C2 to be between 829--1088\,MHz. The measured FWHM width and the polarization fractions for both sub-pulses are in Table~\ref{tab:rm_table_alternate}. The dynamic spectra, along with stokes polarization profile, are shown in Figure \ref{fig:P05_alternate}. In this interpretation of burst structure, we see no significant presence of circular polarization in the sub-pulse C2. The sub-pulse C1 is highly circular polarized ($|V|/I \sim$ 47 per cent), while C2 is $\sim$ 93 per cent linearly polarized. We also find an apparent discrepancy between the best-fitting RM of both sub-pulses, with the difference $\Delta \mathrm{RM} = 6 \pm 3$\RMunits. To confirm this interpretation, we consider the dynamic spectrum for the Stokes V component. We obtain the structure-optimized DM for Stokes V to $418 \pm 1$\DMunits. The absence of a second sub-pulse in the circularly polarized emission also suggests that the Stokes intensity data components are distinct and separate in time. Figure \ref{fig:stokesvplot} shows a comparison of the Stokes V burst structure for the two different DM scenarios.

\begin{table*}
\centering
\normalsize
\begin{threeparttable}
\caption{Measured properties for the two sub-pulses of the Parkes/UWL burst P05 when dedispersed with the individual structure-optimized DM of 418\DMunits. Polarization fractions are obtained after correcting for the best-RM of each sub-pulse. Burst parameters same as Table~\ref{tab:burst_properties}.} 
\label{tab:rm_table_alternate}
\begin{tabular}{lcccccccccc}
\hline
Burst & $\nu_\mathrm{low}$ & $\nu_\mathrm{high}$ & Width & Gaussian & Fluence & S & RM$_\mathrm{nest}$ & P/I & L/I & V/I \\
      & (MHz) & (MHz) & (ms) & S/N & (Jy ms) & (Jy) & \RMunits & & &\\
\hline
C1 & 704  & 1088 &  12.2(3) & 68.1 & 31(1) & 2.5(1) & -619(1) & 0.88(1) & 0.74(1) & $-$0.47(1)\\
C2 & 829  & 1088 &  13(1)   & 17.9 & 13(1) & 0.7(1) & -613(3) & 0.94(4) & 0.93(4) & $-$0.07(3)\\
\hline
\end{tabular}
\end{threeparttable}
\end{table*}

We now attempt to investigate whether the spectral components of the UWL burst P05 are inducing the observed apparent difference in the DM with the other bursts. From the dynamic spectrum (see Figure \ref{fig:P05_alternate}), the burst components appear to have different amounts of dispersion (or drift rates). We identify three distinct components contributing to the burst structure in sub-bands in range 704--829, 829--954, and 954--1088 MHz. We measure the structure-maximized DM to be $423\pm1$, $414\pm1$ and $417\pm2$ \DMunits\,for the three sub-components, respectively. The apparent DM for the brightest component (middle sub-band) of the burst signal agrees with the average global DM of the source. The other components are more consistent with the alternative scenario (DM = 418\DMunits) and thus contribute to the apparent access in DM for the overall P05 burst structure. Specifically, the low-frequency component appears distinct and bifurcated from the bright one. The differences in apparent DM could be evidence that the structure-optimization methods are not suitable for measuring the DM using the entire burst structure. 

Bursts from repeating FRB sources tend to have their dynamic spectrum drifting downwards in frequency \citep{Hessels:2019}. The mechanism responsible for the downward drifting and the presence of multiple burst components embedded into each other is not established yet. It is unclear if the apparent dispersion excess in the burst due to the embedded sub-structures corresponds to any physical mechanism or simply an effect of unresolved burst components \citep{Platts:2021}. Propagation effects such as plasma lensing \citep{Cordes:2017} or the sub-bursts emission along different lines of sight (possibly at different times) through a dense plasma can result in a DM discrepancy within the sub-bursts, but not on a scale of what we observe in the UWL burst P05. The dilemma in deciding the best-DM or the correct alignment of the burst structure creates difficulty in studying the spectrum and its dependence on the observed frequency. To form the spectrum, one needs to determine the burst extent in time and then accumulate the signal contained in those time bins. In the case of drifted bursts, we must trace the signal in the time bins following the drift curve to form the correct spectrum. For the spectral study, we can simply utilize this "extra dispersion" due to drifting and align the structure in time as much as possible to form the correct spectrum for each drifting component. Thus, in the case of UWL burst P05, a DM of 418\DMunits~is more appropriate to study the spectrum of the whole burst structure.

\section{Discussion}\label{sec:discussion}
\subsection{Polarization}
We observe a high polarization fraction, $P/I \sim$ 0.6--0.9 in all five UWL bursts, with similarly high degrees of linear polarization $L/I \sim$ 0.5--0.9. The high polarization is consistent with the published sample of FRBs \citep{Cordes:2019, Fonseca:2020}. However, we measure significant circular polarization (as high as $|V|/I = 0.47$) in two of the UWL repeat bursts (P03 and P05). This source is also the first case of a repeating FRB source to exhibit circularly polarized emission in its bursts \citep{Hilmarsson:2021b}. While several non-repeating FRBs displayed significant circular components, few have circular polarization fractions as large as the UWL burst P05 [e.g., FRB\,20181112A with a maximum $|V|/I = 0.34$ \citep{Cho:2020} and FRB\,20190611B with 0.57 \citep{Day:2020}]. We also find that the linear PA varies across the burst P05 pulse profile, with PA values spanning a range of $\sim$50 deg. Such large variations in PA indicate a change in the magnetic field orientation with respect to the line of sight. One natural explanation is that the emission traverses different paths through a magnetosphere around the progenitor.

While strong circularly polarized emission is found in bursts P03 and P05, it is noticeably absent in the interim burst P04. The three bursts are respectively separated by $\sim$9 min. Hence, the magnetic field or plasma density in the magnetosphere must be varying on an equivalently short time-scale. In the burst P05 itself, while the brighter sub-pulse exhibits a rich presence of circular polarization, in contrast, the other sub-pulse displays none. At the same time, we observe an apparent increase in the RM and degree of linear polarization in the second sub-pulse. Similar distinct polarization properties have been observed in sub-pulses of a few non-repeating FRBs, 20190611B \citep{Day:2020} and 20181112A \citep{Cho:2020}. For FRB\,20181112A, \citet{Cho:2020} concluded that such polarization variations could originate from the propagation of the burst signal through a birefringent medium, likely containing a relativistic plasma. It is quite intriguing that the only pulse with a large linear PA variation in our sample also has a strong presence of circular polarization. On the other hand, all other UWL bursts and the second sub-pulse in P05 show a flat linear PA. Such correlations between PA swings and circular polarization in the pulse profiles have been extensively studied and debated for radio pulsars \citep{Radhakrishnan:1990, Han:1998}. We do not find any transition in the handedness of the circular polarization within the P05 sub-pulse. This symmetric nature of circular polarization observed in pulse profiles of pulsars is usually interpreted due to the propagation effects in the neutron star magnetosphere \citep{Radhakrishnan:1990}.

FRB models involving relativistic shocks far from the progenitor, such as the synchrotron maser model \citep{Metzger:2019, Beloborodov:2020} predict high linear polarization and non-varying PA. These models successfully explain the bursts from FRB\,121102, and many other repeating sources. Nevertheless, the drawback of these models is that they cannot, in their current form, explain the 100 per cent linear polarization \citep{Lyubarsky:2021} observed in some FRBs \citep{Gajjar:2018, Oslowski:2019} or large variations in PA observed in FRB\,20180301A \citep{Luo:2020}. A near-field magnetic re-connection emission model \citep{Lyubarsky:2020} on the other hand, can explain the observed linear polarization up to 100 per cent and variable PA swings between bursts but cannot explain the wide range of luminosities observed in repeating FRBs \citep{Lyubarsky:2021}. Here in our case as well, these prevailing models cannot explain the significant circular polarization fraction and evolving PA observed in burst P05. However, we see such effects in Galactic radio pulsars and magnetars \citep{Noutsos:2009, Dai:2019} and thus, the polarization properties of bursts from FRB20201124A are at least phenomenologically consistent with a magnetospheric origin.

Further development of magnetospheric models is necessary, taking the production of both linearly and circularly polarized components into account. It is also possible that multiple such processes -- intrinsic to the emission and extrinsic propagation effects are responsible to explain the diverse polarization properties of bursts from FRB sources.

\subsection{Burst properties}
FRB20201124A is not only a remarkably bright repeating source; it is also very luminous for a repeating FRB. All of the ASKAP and Parkes repeat bursts reported here are highly energetic for repeating FRB sources, with an inferred isotropic peak luminosity spanning two orders of magnitude in the range $\sim 10^{41}$--$10^{43}$\, erg s$^{-1}$. Two of the ASKAP repeat bursts (A01 and A02), separated by a duration of $\sim$ 37 min, have a fluence > 100 Jy~ms and are clearly in the sample of the brightest repetitions from any FRB source. Burst A02 is one of the most energetic repeat burst from an FRB source detected to date (second only to FRB\,20190711A), with an isotropic energy of $1.2 \times 10^{41}$~erg. This is comparable to those of the localized (cosmological) ASKAP FRBs and inferred of the non-localized ASKAP fly's eye FRBs \citep{Shannon:2018}. Since all of the burst flux is concentrated in the lower 114 MHz part and might extend below the ASKAP observing band, the burst could be even more energetic than determined here. Progenitor models assuming a catastrophic event as the cardinal event and the subsequent bursts from the remaining progenitor as weaker events cannot explain the observed high energy in the repetitions from this source \citep{Jiang:2020}. We already have observed the bursts from this source differing in detected fluence by three orders in magnitude \citep{Hilmarsson:2021b} and considering fainter detections \citep{Xu:2021ATel14518} like other repeating FRBs; this would mean the emitted peak luminosities span five orders of magnitude for bursts from a single object.

Repeating FRBs show a varying degree of repetition with some sources emitting bursts in a periodic modulation of activity window of 16.35 days in FRB20180916B \citep{CHIME:2020_periodicity} and 157 days in FRB20121102A \citep{Rajwade:2020}. While other repeating sources have not shown any sign of such periodicity yet, given the high range of modulation time of these two FRB sources, a conclusive analysis for any source needs years of regular monitoring. The detection of five bright bursts in just 1 hr of follow-up with Parkes/UWL clearly indicates a high burst rate from the source of FRB20201124A when the source is active. The inferred value of cumulative power-law index $\gamma = -0.8\pm0.1$ using all reported observations from this repeating source is in agreement with previous such measurements for FRB20121102A \citep{James:2019_source_count, Cruces:2021}. Using the inferred value of $\beta = -5.8\pm0.8$ for the spectral dependence, we compare the burst rate from this FRB source with the two most prolific repeaters. \citet{Pleunis:2021_LOFAR} measure the CHIME detection rate of $\sim 1.5$ bursts h$^{-1}$ above 5.1 Jy~ms for FRB20180916B in the best active window around each activity cycle. Similarly, the peak burst rate determined for FRB\,20121102A with FAST detections is $\sim 122$ h$^{-1}$ above 0.015 Jy~ms \citep{Li:2021}. We find the burst rate of FRB\,20201124A to be $\sim 9$ and $\sim 28$~h$^{-1}$ after scaling to the CHIME and FAST sensitivities, respectively. Clearly, FRB\,20201124A is one of the most active sources of repeating FRBs.

Bursts from repeating FRB sources exhibit a common feature of downward drift in frequency, where sub-bursts arrival get delayed at lower frequencies \citep{CHIME:2019_8repeaters, Hessels:2019}. Significant evidence of upward drifting with frequency has not yet been observed in repeating FRBs. There have been few instances of visible upward drifting in the burst spectra \citep{Chawla:2020, Day:2020}; however, the burst envelope is found to be relatively large and therefore too complicated to determine if the sub-bursts are part of the same envelope. Some bursts have also displayed complex sub-structures drifting both upwards and downwards in the same envelope \citep{Hilmarsson:2021, Pleunis:2021_LOFAR}. The second component in the UWL burst P05 (see Fig. \ref{fig:p05_uwl}) has a higher central frequency of the spectral envelope than the main pulse suggesting an upward drift. However, there is an ambiguity whether these two sub-bursts are separate and individually drifting downward or are part of the same envelope, suggesting sub-bursts can also drift upward.

The characteristic drifting of burst sub-structure can also result in the dynamic signal spectrum deviating from the usual $\nu^{-2}$ dependence when corrected for the structure-maximized DM. We see a similar case here in the burst P05. When corrected for the average source DM of 413 \DMunits, we see a hint of "bifurcating" structure in the main pulse at $\sim$\,820 MHz, following a different DM. Similar features have been previously observed in some of the bursts from other repeating FRBs sources \citep{Hessels:2019, Marthi:2020, Platts:2021}. The embedding of different sub-burst components is possibly due to the multiple unresolved drifting components in the pulse \citep{Platts:2021}. The complex burst structure can be resolved into at least three distinct components in frequency, each with separate DMs. The best way to explain the whole burst structure and study the spectrum is using the individual structure-optimized DM of 418 \DMunits. The circularly polarized spectra also support this interpretation. The complicated structure of burst P05 with drifting in frequency limits our ability to determine the amount of dispersion with certainty. Whether this apparent ``DM excess'' of 5\DMunits~in burst P05 is real is subject to further modelling using the burst spectro-polarimetric properties. We are limited in any further analysis of the burst structure due to the resolution of available data and the impact of radio frequency interference.

\subsection{Emission properties}
The inferred magnitude of the rotation measure ($|\text{RM}| = 614$\RMunits) for FRB\,20201124A bursts is the second-largest observed in the FRB population.  The Milky Way Faraday rotation along the line of sight of this source is estimated to be $-57$ \RMunits \citep{Oppermann:2015} indicating a significant extra-Galactic component. \citet{Yang:2020} argues for a relation between RM and the luminosity of the PRS associated with an FRB source to explain the high $|\text{RM}| \sim 10^5$\RMunits~measured for FRB20121102A. The non-association of FRB\,20201124A source with a PRS \citep{Marcote:2021ATel14603} and a smaller absolute RM is consistent with their relation. RM variations (between $-624$ and $-579$\RMunits) on a time-scale of $\sim$1~d has been observed in the bursts from this source \citep{Hilmarsson:2021b} similar to another repeating source FRB\,20180301A \citep{Luo:2020}. Although these variations are subject to under-estimation of measurement uncertainties, further high-resolution polarimetric observations of bursts from FRB\,20201124A will strengthen the confidence of measured instability in RM.

FRB\,20201124A is an extreme case in the population of repeating FRBs. The bursts from this source show all characteristic properties (non-varying PA, high linear polarization, large temporal widths, complex morphology and downward-drifting across frequency bands) observed in other repeaters, as well as some unique features. The bursts are incredibly bright for a repeating FRB. Some of the bursts from this source show a high degree of circularly polarized emission and PA swings. Even after all these remarkable properties, the measured RM for this source is not excessively extreme, as in the case of prolific repeater FRB\,20121102A \citep{Michilli:2018}. Another example is FRB\,20180301A, the only other repeating FRB source with PA swings in the burst profiles but with a moderate RM of $\sim$~530~\RMunits \citep{Luo:2020}. This distinction suggests that such extreme repetition and polarization properties do not necessarily imply an extreme environment.

Repeating FRBs have been detected at radio frequencies from 110 MHz \citep{Pleunis:2021_LOFAR} up to 8 GHz \citep{Gajjar:2018}. Multiple attempts have been made to monitor the repeating FRB sources simultaneously at different frequencies using different radio telescopes \citep{Law:2017, Majid:2020} to characterize the FRB spectra. Only two successful concurrent detections at different radio frequencies have been found; a repeat burst from the FRB\,20121102A source in the frequency range 1.2--1.7 and 2.5--3.5 GHz \citep{Law:2017} and other from the FRB\,20180916B source in the range 300--500 MHz \citep{Chawla:2020}. The repeating source FRB\,20180916B displays frequency-dependent activity with bursts at higher frequency arriving at the start of the activity window, indicating a correlation between observing frequency and phase of the burst emission \citep{Aggarwal:2020RNAAS, Marazuela:2021, Pleunis:2021_LOFAR}. In the other active repeating source FRB\,20121102A, \citet{Gourdji:2019} found that the preferred emission frequencies of repeat bursts change on a time-scale of $\sim$days. Wide-band monitoring of repeat bursts using a single instrument like the Parkes/UWL can better constrain the spectral extent and its evolution with time.

We observe a similar change of the emission frequencies in the bursts from FRB\,20201124A. A timeline of all the bursts detected between 2021 April 1 and 8 with their spectral envelope is shown in Figure~\ref{fig:timeline}. In the hour-long observation with the Parkes/UWL system, all the bursts detected are band-limited with $\Delta \nu / \nu \sim 0.09$--0.4. Interestingly, the ASKAP burst A08 detected $\approx 1$\,h after UWL burst P05 has a comparable fluence but $\Delta \nu / \nu \sim 0.2$. We see the emission spectral coverage of bursts switching to a different band on a time-scale of $\sim$minutes. This "band-switching" assumes that repeat bursts are narrow-band \citep{Pleunis:2021} and there is no signal beyond the observed spectral envelope. Even in a scenario where emission extends beyond the observed envelope and below our detection threshold, the centre of the emission envelope switches with time. We also see multiple bursts at $\sim$\,600 MHz reported by CHIME and uGMRT \citep{Marthi:2021} around our observing session on April 5. Although our frequency coverage extends only to 704 MHz, assuming the bursts are narrow-band, this would suggest the switch of peak emission frequency extends from 400 to 1500 MHz within a few hours. A clear evolution of spectral envelope in frequency in this time frame can also be interpreted. Determination of a periodic modulation in this FRB source, if any, will enable us to study the frequency-phase correlation, which will decide if this "band-switching" is intrinsic to the burst emission or propagation effects like plasma lensing are at play.

\section{Conclusions}\label{sec:conclusion}
We have presented a set of 16 repeat bursts from the source of CHIME-discovered FRB\,20201124A using ASKAP and Parkes. We found that all the Parkes/UWL bursts are band-limited and spectrally confined in the instrument's frequency range. They show no evidence of emission in more than 3 GHz of bandwidth. ASKAP bursts also appear to be narrow-band, with most of them extending beyond one of the band edges.  Here, we showed the detection of a band-limited burst with $\sim$47 per cent circular polarization, a first such case in a repeating FRB source. We suggest that the complex burst structure in this burst can be interpreted using a DM excess from the average DM of the source for spectral studies. The contrasting polarimetric properties in the two sub-pulses of this burst indicate that multiple mechanisms (intrinsic and propagation-induced) might be responsible for the burst emission. The variety of observed pulse profiles of the Parkes/UWL repeat bursts along with varying polarization features within an hour of observation suggests a dynamic and evolving magnetic environment around the FRB progenitor.

The observed diversity in polarimetric properties, either the stability of PA \citep{Luo:2020} or the presence of a circular component \citep{Dai:2021} clearly disapprove its use in discriminating against the different sub-classes (repeaters and non-repeaters) of FRBs. Additionally, we found that the preferred frequency of burst emission evolves on a time-scale of a few hours. These detections further highlight the importance of sub-band searches for FRBs with wideband instruments. Thorough coverage of repeating FRBs with more such multi-observatory campaigns will allow us to study the spectral evolution in detail.

\section*{Acknowledgements}
We would like to thank C.~W.~James, J.~X.~Prochaska and N.~Thyagarajan for careful reading of the manuscript and useful comments.
PK acknowledge support through the Australian Research Council (ARC) grant FL150100148. 
RMS acknowledges support through ARC grants  DP180100857 and FT190100155. 
ATD is the recipient of an ARC Future Fellowship (FT150100415).
This work was performed on the OzSTAR national facility at Swinburne University of Technology. The OzSTAR program receives funding in part from the Astronomy National Collaborative Research Infrastructure Strategy (NCRIS) allocation provided by the Australian Government.
The Parkes Radio Telescope (\textit{Murriyang}) and the Australian Square Kilometre Array Pathfinder are part of the Australia Telescope National Facility (grid.421683.a) which is managed by CSIRO. Operation of ASKAP is funded by the Australian Government with support from the NCRIS. ASKAP uses the resources of the Pawsey Supercomputing Centre. Establishment of ASKAP, the Murchison Radio-astronomy Observatory (MRO) and the Pawsey Supercomputing Centre are initiatives of the Australian Government, with support from the Government of Western Australia and the Science and Industry Endowment Fund. We acknowledge the Wajarri Yamatji people as the traditional owners of the MRO site and the Wiradjuri people as the traditional owners of the Parkes observatory site.
This research has made use of NASA's Astrophysics Data System Bibliographic Services, Astronomer's Telegram and software packages, including: \software{MATPLOTLIB} \citep{Hunter:2007_matplotlib}, \software{ASTROPY} \citep{Astropy:2013, Astropy:2018}, \software{NUMPY} \citep{Harris:2020_numpy}, \software{LMFIT} \citep{Newville:2016}, \software{PyMultiNest} \citep{Buchner:2014}, \software{YOUR} \citep{Aggarwal:2020JOSS} and \software{CMASHER} for colormaps \citep{Velden:2020_cmasher}.

\section*{Data Availability}

The data underlying this article will be shared on reasonable request to the corresponding author.



\bibliographystyle{mnras}
\bibliography{references} 

\begin{thebibliography}{}
\makeatletter
\relax
\def\mn@urlcharsother{\let\do\@makeother \do\$\do\&\do\#\do\^\do\_\do\%\do\~}
\def\mn@doi{\begingroup\mn@urlcharsother \@ifnextchar [ {\mn@doi@}
  {\mn@doi@[]}}
\def\mn@doi@[#1]#2{\def\@tempa{#1}\ifx\@tempa\@empty \href
  {http://dx.doi.org/#2} {doi:#2}\else \href {http://dx.doi.org/#2} {#1}\fi
  \endgroup}
\def\mn@eprint#1#2{\mn@eprint@#1:#2::\@nil}
\def\mn@eprint@arXiv#1{\href {http://arxiv.org/abs/#1} {{\tt arXiv:#1}}}
\def\mn@eprint@dblp#1{\href {http://dblp.uni-trier.de/rec/bibtex/#1.xml}
  {dblp:#1}}
\def\mn@eprint@#1:#2:#3:#4\@nil{\def\@tempa {#1}\def\@tempb {#2}\def\@tempc
  {#3}\ifx \@tempc \@empty \let \@tempc \@tempb \let \@tempb \@tempa \fi \ifx
  \@tempb \@empty \def\@tempb {arXiv}\fi \@ifundefined
  {mn@eprint@\@tempb}{\@tempb:\@tempc}{\expandafter \expandafter \csname
  mn@eprint@\@tempb\endcsname \expandafter{\@tempc}}}

\bibitem[\protect\citeauthoryear{{Aggarwal}, {Law}, {Burke-Spolaor}, {Bower},
  {Butler}, {Demorest}, {Linford}  \& {Lazio}}{{Aggarwal}
  et~al.}{2020a}]{Aggarwal:2020RNAAS}
{Aggarwal} K.,  {Law} C.~J.,  {Burke-Spolaor} S.,  {Bower} G.,  {Butler} B.~J.,
   {Demorest} P.,  {Linford} J.,   {Lazio} T.~J.~W.,  2020a, \mn@doi [Research
  Notes of the {AAS}] {10.3847/2515-5172/ab9f33}, \href
  {https://ui.adsabs.harvard.edu/abs/2020RNAAS...4...94A} {4, 94}

\bibitem[\protect\citeauthoryear{{Aggarwal} et~al.,}{{Aggarwal}
  et~al.}{2020b}]{Aggarwal:2020JOSS}
{Aggarwal} K.,  et~al., 2020b, \mn@doi [The Journal of Open Source Software]
  {10.21105/joss.02750}, \href
  {https://ui.adsabs.harvard.edu/abs/2020JOSS....5.2750A} {5, 2750}

\bibitem[\protect\citeauthoryear{{Ashton} et~al.,}{{Ashton}
  et~al.}{2019}]{Ashton:2019}
{Ashton} G.,  et~al., 2019, \mn@doi [\apjs] {10.3847/1538-4365/ab06fc}, \href
  {https://ui.adsabs.harvard.edu/abs/2019ApJS..241...27A} {241, 27}

\bibitem[\protect\citeauthoryear{{Astropy Collaboration} et~al.,}{{Astropy
  Collaboration} et~al.}{2013}]{Astropy:2013}
{Astropy Collaboration} et~al., 2013, \mn@doi [\aap]
  {10.1051/0004-6361/201322068}, \href
  {https://ui.adsabs.harvard.edu/abs/2013A&A...558A..33A} {558, A33}

\bibitem[\protect\citeauthoryear{{Astropy Collaboration} et~al.,}{{Astropy
  Collaboration} et~al.}{2018}]{Astropy:2018}
{Astropy Collaboration} et~al., 2018, \mn@doi [\aj] {10.3847/1538-3881/aabc4f},
  \href {https://ui.adsabs.harvard.edu/abs/2018AJ....156..123A} {156, 123}

\bibitem[\protect\citeauthoryear{{Bannister} et~al.,}{{Bannister}
  et~al.}{2017}]{Bannister:2017}
{Bannister} K.~W.,  et~al., 2017, \mn@doi [\apjl] {10.3847/2041-8213/aa71ff},
  \href {https://ui.adsabs.harvard.edu/abs/2017ApJ...841L..12B} {841, L12}

\bibitem[\protect\citeauthoryear{{Bannister}, {Zackay}, {Qiu}, {James}  \&
  {Shannon}}{{Bannister} et~al.}{2019a}]{fredda_ascl}
{Bannister} K.,  {Zackay} B.,  {Qiu} H.,  {James} C.,   {Shannon} R.,  2019a,
  {FREDDA: A fast, real-time engine for de-dispersing amplitudes} (\mn@eprint
  {ascl} {1906.003})

\bibitem[\protect\citeauthoryear{{Bannister} et~al.,}{{Bannister}
  et~al.}{2019b}]{Bannister:2019_localization}
{Bannister} K.~W.,  et~al., 2019b, \mn@doi [Science] {10.1126/science.aaw5903},
  \href {https://ui.adsabs.harvard.edu/abs/2019Sci...365..565B} {365, 565}

\bibitem[\protect\citeauthoryear{{Barsdell}}{{Barsdell}}{2012}]{Barsdell:2012PhDT}
{Barsdell} B.~R.,  2012, PhD thesis, Swinburne University of Technology, \url
  {http://hdl.handle.net/1959.3/313933}

\bibitem[\protect\citeauthoryear{{Beloborodov}}{{Beloborodov}}{2020}]{Beloborodov:2020}
{Beloborodov} A.~M.,  2020, \mn@doi [\apj] {10.3847/1538-4357/ab83eb}, \href
  {https://ui.adsabs.harvard.edu/abs/2020ApJ...896..142B} {896, 142}

\bibitem[\protect\citeauthoryear{{Bhandari} et~al.,}{{Bhandari}
  et~al.}{2018}]{Bhandari:2018}
{Bhandari} S.,  et~al., 2018, \mn@doi [\mnras] {10.1093/mnras/stx3074}, \href
  {https://ui.adsabs.harvard.edu/abs/2018MNRAS.475.1427B} {475, 1427}

\bibitem[\protect\citeauthoryear{{Bhandari} et~al.,}{{Bhandari}
  et~al.}{2020}]{Bhandari:2020}
{Bhandari} S.,  et~al., 2020, \mn@doi [\apjl] {10.3847/2041-8213/ab672e}, \href
  {https://ui.adsabs.harvard.edu/abs/2020ApJ...895L..37B} {895, L37}

\bibitem[\protect\citeauthoryear{{Bhardwaj} et~al.,}{{Bhardwaj}
  et~al.}{2021}]{Bhardwaj:2021}
{Bhardwaj} M.,  et~al., 2021, \mn@doi [\apjl] {10.3847/2041-8213/abeaa6}, \href
  {https://ui.adsabs.harvard.edu/abs/2021ApJ...910L..18B} {910, L18}

\bibitem[\protect\citeauthoryear{{Bochenek}, {Ravi}  \& {Dong}}{{Bochenek}
  et~al.}{2021}]{Bochenek:2021}
{Bochenek} C.~D.,  {Ravi} V.,   {Dong} D.,  2021, \mn@doi [\apjl]
  {10.3847/2041-8213/abd634}, \href
  {https://ui.adsabs.harvard.edu/abs/2021ApJ...907L..31B} {907, L31}

\bibitem[\protect\citeauthoryear{{Buchner} et~al.,}{{Buchner}
  et~al.}{2014}]{Buchner:2014}
{Buchner} J.,  et~al., 2014, \mn@doi [\aap] {10.1051/0004-6361/201322971},
  \href {https://ui.adsabs.harvard.edu/abs/2014A&A...564A.125B} {564, A125}

\bibitem[\protect\citeauthoryear{{CHIME/FRB Collaboration}}{{CHIME/FRB
  Collaboration}}{2021}]{CHIME:2021ATel14497}
{CHIME/FRB Collaboration} 2021, The Astronomer's Telegram, \href
  {https://ui.adsabs.harvard.edu/abs/2021ATel14497....1C} {14497, 1}

\bibitem[\protect\citeauthoryear{{CHIME/FRB Collaboration} et~al.,}{{CHIME/FRB
  Collaboration} et~al.}{2019a}]{CHIME:2019}
{CHIME/FRB Collaboration} et~al., 2019a, \mn@doi [\nat]
  {10.1038/s41586-018-0864-x}, \href
  {https://ui.adsabs.harvard.edu/abs/2019Natur.566..235C} {566, 235}

\bibitem[\protect\citeauthoryear{{CHIME/FRB Collaboration} et~al.,}{{CHIME/FRB
  Collaboration} et~al.}{2019b}]{CHIME:2019_8repeaters}
{CHIME/FRB Collaboration} et~al., 2019b, \mn@doi [\apjl]
  {10.3847/2041-8213/ab4a80}, \href
  {https://ui.adsabs.harvard.edu/abs/2019ApJ...885L..24C} {885, L24}

\bibitem[\protect\citeauthoryear{{CHIME/FRB Collaboration} et~al.,}{{CHIME/FRB
  Collaboration} et~al.}{2020}]{CHIME:2020_periodicity}
{CHIME/FRB Collaboration} et~al., 2020, \mn@doi [\nat]
  {10.1038/s41586-020-2398-2}, \href
  {https://ui.adsabs.harvard.edu/abs/2020arXiv200110275T} {582, 351}

\bibitem[\protect\citeauthoryear{{CHIME/FRB Collaboration} et~al.,}{{CHIME/FRB
  Collaboration} et~al.}{2021}]{CHIME:2021}
{CHIME/FRB Collaboration} et~al., 2021, \mn@doi [\apjs]
  {10.3847/1538-4365/ac33ab}, \href
  {https://ui.adsabs.harvard.edu/abs/2021ApJS..257...59A} {257, 59}

\bibitem[\protect\citeauthoryear{{Caleb} et~al.,}{{Caleb}
  et~al.}{2018}]{Caleb:2018}
{Caleb} M.,  et~al., 2018, \mn@doi [\mnras] {10.1093/mnras/sty1137}, \href
  {https://ui.adsabs.harvard.edu/abs/2018MNRAS.478.2046C} {478, 2046}

\bibitem[\protect\citeauthoryear{{Campana}}{{Campana}}{2021}]{Campana:2021ATel14523}
{Campana} S.,  2021, The Astronomer's Telegram, \href
  {https://ui.adsabs.harvard.edu/abs/2021ATel14523....1C} {14523, 1}

\bibitem[\protect\citeauthoryear{{Chatterjee} et~al.,}{{Chatterjee}
  et~al.}{2017}]{Chatterjee:2017}
{Chatterjee} S.,  et~al., 2017, \mn@doi [\nat] {10.1038/nature20797}, \href
  {https://ui.adsabs.harvard.edu/abs/2017Natur.541...58C} {541, 58}

\bibitem[\protect\citeauthoryear{{Chawla} et~al.,}{{Chawla}
  et~al.}{2020}]{Chawla:2020}
{Chawla} P.,  et~al., 2020, \mn@doi [\apjl] {10.3847/2041-8213/ab96bf}, \href
  {https://ui.adsabs.harvard.edu/abs/2020ApJ...896L..41C} {896, L41}

\bibitem[\protect\citeauthoryear{{Cho} et~al.,}{{Cho} et~al.}{2020}]{Cho:2020}
{Cho} H.,  et~al., 2020, \mn@doi [\apjl] {10.3847/2041-8213/ab7824}, \href
  {https://ui.adsabs.harvard.edu/abs/2020ApJ...891L..38C} {891, L38}

\bibitem[\protect\citeauthoryear{{Cordes} \& {Chatterjee}}{{Cordes} \&
  {Chatterjee}}{2019}]{Cordes:2019}
{Cordes} J.~M.,  {Chatterjee} S.,  2019, \mn@doi [\araa]
  {10.1146/annurev-astro-091918-104501}, \href
  {https://ui.adsabs.harvard.edu/abs/2019ARA&A..57..417C} {57, 417}

\bibitem[\protect\citeauthoryear{{Cordes} \& {McLaughlin}}{{Cordes} \&
  {McLaughlin}}{2003}]{Cordes:2003}
{Cordes} J.~M.,  {McLaughlin} M.~A.,  2003, \mn@doi [\apj] {10.1086/378231},
  \href {https://ui.adsabs.harvard.edu/abs/2003ApJ...596.1142C} {596, 1142}

\bibitem[\protect\citeauthoryear{{Cordes}, {Wasserman}, {Hessels}, {Lazio},
  {Chatterjee}  \& {Wharton}}{{Cordes} et~al.}{2017}]{Cordes:2017}
{Cordes} J.~M.,  {Wasserman} I.,  {Hessels} J.~W.~T.,  {Lazio} T.~J.~W.,
  {Chatterjee} S.,   {Wharton} R.~S.,  2017, \mn@doi [\apj]
  {10.3847/1538-4357/aa74da}, \href
  {https://ui.adsabs.harvard.edu/abs/2017ApJ...842...35C} {842, 35}

\bibitem[\protect\citeauthoryear{{Cruces} et~al.,}{{Cruces}
  et~al.}{2021}]{Cruces:2021}
{Cruces} M.,  et~al., 2021, \mn@doi [\mnras] {10.1093/mnras/staa3223}, \href
  {https://ui.adsabs.harvard.edu/abs/2021MNRAS.500..448C} {500, 448}

\bibitem[\protect\citeauthoryear{{Dai} et~al.,}{{Dai} et~al.}{2019}]{Dai:2019}
{Dai} S.,  et~al., 2019, \mn@doi [\apjl] {10.3847/2041-8213/ab0e7a}, \href
  {https://ui.adsabs.harvard.edu/abs/2019ApJ...874L..14D} {874, L14}

\bibitem[\protect\citeauthoryear{{Dai} et~al.,}{{Dai} et~al.}{2021}]{Dai:2021}
{Dai} S.,  et~al., 2021, \mn@doi [\apj] {10.3847/1538-4357/ac193d}, \href
  {https://ui.adsabs.harvard.edu/abs/2021ApJ...920...46D} {920, 46}

\bibitem[\protect\citeauthoryear{{Day} et~al.,}{{Day} et~al.}{2020}]{Day:2020}
{Day} C.~K.,  et~al., 2020, \mn@doi [\mnras] {10.1093/mnras/staa2138}, \href
  {https://ui.adsabs.harvard.edu/abs/2020MNRAS.497.3335D} {497, 3335}

\bibitem[\protect\citeauthoryear{{Day}, {Bhandari}, {Deller}, {Shannon}  \&
  {Moss}}{{Day} et~al.}{2021a}]{Day:2021ATel14515}
{Day} C.~K.,  {Bhandari} S.,  {Deller} A.~T.,  {Shannon} R.~M.,   {Moss} V.~A.,
   2021a, The Astronomer's Telegram, \href
  {https://ui.adsabs.harvard.edu/abs/2021ATel14515....1D} {14515, 1}

\bibitem[\protect\citeauthoryear{{Day}, {Bhandari}, {Deller}, {Shannon}  \&
  {ASKAP-CRAFT Survey Science Project}}{{Day}
  et~al.}{2021b}]{Day:2021ATel14592}
{Day} C.~K.,  {Bhandari} S.,  {Deller} A.~T.,  {Shannon} R.~M.,   {ASKAP-CRAFT
  Survey Science Project} 2021b, The Astronomer's Telegram, \href
  {https://ui.adsabs.harvard.edu/abs/2021ATel14592....1D} {14592, 1}

\bibitem[\protect\citeauthoryear{{Everett} \& {Weisberg}}{{Everett} \&
  {Weisberg}}{2001}]{Everett:2001}
{Everett} J.~E.,  {Weisberg} J.~M.,  2001, \mn@doi [\apj] {10.1086/320652},
  \href {https://ui.adsabs.harvard.edu/abs/2001ApJ...553..341E} {553, 341}

\bibitem[\protect\citeauthoryear{{Farah} et~al.,}{{Farah}
  et~al.}{2021}]{Farah:2021ATel14676}
{Farah} W.,  et~al., 2021, The Astronomer's Telegram, \href
  {https://ui.adsabs.harvard.edu/abs/2021ATel14676....1F} {14676, 1}

\bibitem[\protect\citeauthoryear{{Feroz}, {Hobson}  \& {Bridges}}{{Feroz}
  et~al.}{2009}]{Feroz:2009}
{Feroz} F.,  {Hobson} M.~P.,   {Bridges} M.,  2009, \mn@doi [\mnras]
  {10.1111/j.1365-2966.2009.14548.x}, \href
  {https://ui.adsabs.harvard.edu/abs/2009MNRAS.398.1601F} {398, 1601}

\bibitem[\protect\citeauthoryear{{Fong} et~al.,}{{Fong}
  et~al.}{2021}]{Fong:2021}
{Fong} W.-f.,  et~al., 2021, \mn@doi [\apjl] {10.3847/2041-8213/ac242b}, \href
  {https://ui.adsabs.harvard.edu/abs/2021ApJ...919L..23F} {919, L23}

\bibitem[\protect\citeauthoryear{{Fonseca} et~al.,}{{Fonseca}
  et~al.}{2020}]{Fonseca:2020}
{Fonseca} E.,  et~al., 2020, \mn@doi [\apjl] {10.3847/2041-8213/ab7208}, \href
  {https://ui.adsabs.harvard.edu/abs/2020ApJ...891L...6F} {891, L6}

\bibitem[\protect\citeauthoryear{{Gajjar} et~al.,}{{Gajjar}
  et~al.}{2018}]{Gajjar:2018}
{Gajjar} V.,  et~al., 2018, \mn@doi [\apj] {10.3847/1538-4357/aad005}, \href
  {https://ui.adsabs.harvard.edu/abs/2018ApJ...863....2G} {863, 2}

\bibitem[\protect\citeauthoryear{{Gourdji}, {Michilli}, {Spitler}, {Hessels},
  {Seymour}, {Cordes}  \& {Chatterjee}}{{Gourdji} et~al.}{2019}]{Gourdji:2019}
{Gourdji} K.,  {Michilli} D.,  {Spitler} L.~G.,  {Hessels} J.~W.~T.,  {Seymour}
  A.,  {Cordes} J.~M.,   {Chatterjee} S.,  2019, \mn@doi [\apjl]
  {10.3847/2041-8213/ab1f8a}, \href
  {https://ui.adsabs.harvard.edu/abs/2019ApJ...877L..19G} {877, L19}

\bibitem[\protect\citeauthoryear{{Han}, {Manchester}, {Xu}  \& {Qiao}}{{Han}
  et~al.}{1998}]{Han:1998}
{Han} J.~L.,  {Manchester} R.~N.,  {Xu} R.~X.,   {Qiao} G.~J.,  1998, \mn@doi
  [\mnras] {10.1046/j.1365-8711.1998.01869.x}, \href
  {https://ui.adsabs.harvard.edu/abs/1998MNRAS.300..373H} {300, 373}

\bibitem[\protect\citeauthoryear{{Harris} et~al.,}{{Harris}
  et~al.}{2020}]{Harris:2020_numpy}
{Harris} C.~R.,  et~al., 2020, \mn@doi [\nat] {10.1038/s41586-020-2649-2},
  \href
  {https://ui-adsabs-harvard-edu.ezproxy.lib.swin.edu.au/abs/2020Natur.585..357H}
  {585, 357}

\bibitem[\protect\citeauthoryear{{Hassall} et~al.,}{{Hassall}
  et~al.}{2012}]{Hassall:2012}
{Hassall} T.~E.,  et~al., 2012, \mn@doi [\aap] {10.1051/0004-6361/201218970},
  \href {https://ui.adsabs.harvard.edu/abs/2012A&A...543A..66H} {543, A66}

\bibitem[\protect\citeauthoryear{{Heintz} et~al.,}{{Heintz}
  et~al.}{2020}]{Heintz:2020}
{Heintz} K.~E.,  et~al., 2020, \mn@doi [\apj] {10.3847/1538-4357/abb6fb}, \href
  {https://ui.adsabs.harvard.edu/abs/2020ApJ...903..152H} {903, 152}

\bibitem[\protect\citeauthoryear{{Herrmann}}{{Herrmann}}{2021}]{Herrmann:2021ATel14556}
{Herrmann} W.,  2021, The Astronomer's Telegram, \href
  {https://ui.adsabs.harvard.edu/abs/2021ATel14556....1H} {14556, 1}

\bibitem[\protect\citeauthoryear{{Hessels} et~al.,}{{Hessels}
  et~al.}{2019}]{Hessels:2019}
{Hessels} J.~W.~T.,  et~al., 2019, \mn@doi [\apjl] {10.3847/2041-8213/ab13ae},
  \href {https://ui.adsabs.harvard.edu/abs/2019ApJ...876L..23H} {876, L23}

\bibitem[\protect\citeauthoryear{{Hilmarsson}, {Spitler}, {Main}  \&
  {Li}}{{Hilmarsson} et~al.}{2021a}]{Hilmarsson:2021b}
{Hilmarsson} G.~H.,  {Spitler} L.~G.,  {Main} R.~A.,   {Li} D.~Z.,  2021a,
  \mn@doi [\mnras] {10.1093/mnras/stab2936}, \href
  {https://ui.adsabs.harvard.edu/abs/2021MNRAS.508.5354H} {508, 5354}

\bibitem[\protect\citeauthoryear{{Hilmarsson} et~al.,}{{Hilmarsson}
  et~al.}{2021b}]{Hilmarsson:2021}
{Hilmarsson} G.~H.,  et~al., 2021b, \mn@doi [\apjl] {10.3847/2041-8213/abdec0},
  \href {https://ui.adsabs.harvard.edu/abs/2021ApJ...908L..10H} {908, L10}

\bibitem[\protect\citeauthoryear{{Hobbs} et~al.,}{{Hobbs}
  et~al.}{2020}]{Hobbs:2020}
{Hobbs} G.,  et~al., 2020, \mn@doi [\pasa] {10.1017/pasa.2020.2}, \href
  {https://ui.adsabs.harvard.edu/abs/2020PASA...37...12H} {37, e012}

\bibitem[\protect\citeauthoryear{{Hotan}, {van Straten}  \&
  {Manchester}}{{Hotan} et~al.}{2004}]{psrchive}
{Hotan} A.~W.,  {van Straten} W.,   {Manchester} R.~N.,  2004, \mn@doi [\pasa]
  {10.1071/AS04022}, \href
  {https://ui.adsabs.harvard.edu/abs/2004PASA...21..302H} {21, 302}

\bibitem[\protect\citeauthoryear{{Hotan} et~al.,}{{Hotan}
  et~al.}{2021}]{Hotan:2021}
{Hotan} A.~W.,  et~al., 2021, \mn@doi [\pasa] {10.1017/pasa.2021.1}, \href
  {https://ui.adsabs.harvard.edu/abs/2021PASA...38....9H} {38, e009}

\bibitem[\protect\citeauthoryear{{Hunter}}{{Hunter}}{2007}]{Hunter:2007_matplotlib}
{Hunter} J.~D.,  2007, \mn@doi [Computing in Science and Engineering]
  {10.1109/MCSE.2007.55}, \href
  {https://ui-adsabs-harvard-edu.ezproxy.lib.swin.edu.au/abs/2007CSE.....9...90H}
  {9, 90}

\bibitem[\protect\citeauthoryear{{James}, {Ekers}, {Macquart}, {Bannister}  \&
  {Shannon}}{{James} et~al.}{2019}]{James:2019_source_count}
{James} C.~W.,  {Ekers} R.~D.,  {Macquart} J.~P.,  {Bannister} K.~W.,
  {Shannon} R.~M.,  2019, \mn@doi [\mnras] {10.1093/mnras/sty3031}, \href
  {https://ui.adsabs.harvard.edu/abs/2019MNRAS.483.1342J} {483, 1342}

\bibitem[\protect\citeauthoryear{{Jiang}, {Wang}, {Luo}, {Du}, {Chen}, {Lee}
  \& {Xu}}{{Jiang} et~al.}{2020}]{Jiang:2020}
{Jiang} J.-C.,  {Wang} W.-Y.,  {Luo} R.,  {Du} S.,  {Chen} X.,  {Lee} K.-J.,
  {Xu} R.-X.,  2020, \mn@doi [Research in Astronomy and Astrophysics]
  {10.1088/1674-4527/20/4/56}, \href
  {https://ui.adsabs.harvard.edu/abs/2020RAA....20...56J} {20, 056}

\bibitem[\protect\citeauthoryear{{Keane} et~al.,}{{Keane}
  et~al.}{2016}]{Keane:2016}
{Keane} E.~F.,  et~al., 2016, \mn@doi [\nat] {10.1038/nature17140}, \href
  {https://ui.adsabs.harvard.edu/abs/2016Natur.530..453K} {530, 453}

\bibitem[\protect\citeauthoryear{{Kilpatrick}, {Fong}, {Prochaska}, {Tejos},
  {Bhandari}  \& {Day}}{{Kilpatrick} et~al.}{2021}]{Kilpatrick:2021ATel14516}
{Kilpatrick} C.~D.,  {Fong} W.,  {Prochaska} J.~X.,  {Tejos} N.,  {Bhandari}
  S.,   {Day} C.~K.,  2021, The Astronomer's Telegram, \href
  {https://ui.adsabs.harvard.edu/abs/2021ATel14516....1K} {14516, 1}

\bibitem[\protect\citeauthoryear{{Kirsten} et~al.,}{{Kirsten}
  et~al.}{2021}]{Kirsten:2021ATel14605}
{Kirsten} F.,  et~al., 2021, The Astronomer's Telegram, \href
  {https://ui.adsabs.harvard.edu/abs/2021ATel14605....1K} {14605, 1}

\bibitem[\protect\citeauthoryear{{Kirsten} et~al.,}{{Kirsten}
  et~al.}{2022}]{Kirsten:2022}
{Kirsten} F.,  et~al., 2022, \mn@doi [\nat] {10.1038/s41586-021-04354-w}, \href
  {https://ui.adsabs.harvard.edu/abs/2022Natur.602..585K} {602, 585}

\bibitem[\protect\citeauthoryear{{Kumar} et~al.,}{{Kumar}
  et~al.}{2019}]{Kumar:2019}
{Kumar} P.,  et~al., 2019, \mn@doi [\apjl] {10.3847/2041-8213/ab5b08}, \href
  {https://ui.adsabs.harvard.edu/abs/2019ApJ...887L..30K} {887, L30}

\bibitem[\protect\citeauthoryear{{Kumar} et~al.,}{{Kumar}
  et~al.}{2021a}]{Kumar:2021}
{Kumar} P.,  et~al., 2021a, \mn@doi [\mnras] {10.1093/mnras/staa3436}, \href
  {https://ui.adsabs.harvard.edu/abs/2021MNRAS.500.2525K} {500, 2525}

\bibitem[\protect\citeauthoryear{{Kumar}, {Shannon}, {Moss}, {Qiu}  \&
  {Bhandari}}{{Kumar} et~al.}{2021b}]{Kumar:2021ATel14502}
{Kumar} P.,  {Shannon} R.~M.,  {Moss} V.,  {Qiu} H.,   {Bhandari} S.,  2021b,
  The Astronomer's Telegram, \href
  {https://ui.adsabs.harvard.edu/abs/2021ATel14502....1K} {14502, 1}

\bibitem[\protect\citeauthoryear{{Kumar}, {Shannon}, {Keane}, {Moss}  \&
  {ASKAP-CRAFT Survey Science Project}}{{Kumar}
  et~al.}{2021c}]{Kumar:2021ATel14508}
{Kumar} P.,  {Shannon} R.~M.,  {Keane} E.,  {Moss} V.~A.,   {ASKAP-CRAFT Survey
  Science Project} 2021c, The Astronomer's Telegram, \href
  {https://ui.adsabs.harvard.edu/abs/2021ATel14508....1K} {14508, 1}

\bibitem[\protect\citeauthoryear{{Lanman} et~al.,}{{Lanman}
  et~al.}{2022}]{Lanman:2022}
{Lanman} A.~E.,  et~al., 2022, \mn@doi [\apj] {10.3847/1538-4357/ac4bc7}, \href
  {https://ui.adsabs.harvard.edu/abs/2022ApJ...927...59L} {927, 59}

\bibitem[\protect\citeauthoryear{{Law} et~al.,}{{Law} et~al.}{2017}]{Law:2017}
{Law} C.~J.,  et~al., 2017, \mn@doi [\apj] {10.3847/1538-4357/aa9700}, \href
  {https://ui.adsabs.harvard.edu/abs/2017ApJ...850...76L} {850, 76}

\bibitem[\protect\citeauthoryear{{Law} et~al.,}{{Law} et~al.}{2018}]{Law:2018}
{Law} C.~J.,  et~al., 2018, \mn@doi [\apjs] {10.3847/1538-4365/aab77b}, \href
  {https://ui.adsabs.harvard.edu/abs/2018ApJS..236....8L} {236, 8}

\bibitem[\protect\citeauthoryear{{Law}, {Tendulkar}, {Clarke}, {Aggarwal}  \&
  {Bethapudy}}{{Law} et~al.}{2021}]{Law:2021ATel14526}
{Law} C.,  {Tendulkar} S.,  {Clarke} T.,  {Aggarwal} K.,   {Bethapudy} S.,
  2021, The Astronomer's Telegram, \href
  {https://ui.adsabs.harvard.edu/abs/2021ATel14526....1L} {14526, 1}

\bibitem[\protect\citeauthoryear{{Li} et~al.,}{{Li} et~al.}{2021}]{Li:2021}
{Li} D.,  et~al., 2021, \mn@doi [\nat] {10.1038/s41586-021-03878-5}, \href
  {https://ui.adsabs.harvard.edu/abs/2021Natur.598..267L} {598, 267}

\bibitem[\protect\citeauthoryear{{Lorimer}}{{Lorimer}}{2011}]{Lorimer:2011_ascl}
{Lorimer} D.~R.,  2011, {SIGPROC: Pulsar Signal Processing Programs}
  (\mn@eprint {ascl} {1107.016})

\bibitem[\protect\citeauthoryear{{Lower}, {Shannon}, {Johnston}  \&
  {Bailes}}{{Lower} et~al.}{2020}]{Lower:2020}
{Lower} M.~E.,  {Shannon} R.~M.,  {Johnston} S.,   {Bailes} M.,  2020, \mn@doi
  [\apjl] {10.3847/2041-8213/ab9898}, \href
  {https://ui.adsabs.harvard.edu/abs/2020ApJ...896L..37L} {896, L37}

\bibitem[\protect\citeauthoryear{{Luo} et~al.,}{{Luo} et~al.}{2020}]{Luo:2020}
{Luo} R.,  et~al., 2020, \mn@doi [\nat] {10.1038/s41586-020-2827-2}, \href
  {https://ui.adsabs.harvard.edu/abs/2020Natur.586..693L} {586, 693}

\bibitem[\protect\citeauthoryear{{Lyubarsky}}{{Lyubarsky}}{2020}]{Lyubarsky:2020}
{Lyubarsky} Y.,  2020, \mn@doi [\apj] {10.3847/1538-4357/ab97b5}, \href
  {https://ui.adsabs.harvard.edu/abs/2020ApJ...897....1L} {897, 1}

\bibitem[\protect\citeauthoryear{{Lyubarsky}}{{Lyubarsky}}{2021}]{Lyubarsky:2021}
{Lyubarsky} Y.,  2021, \mn@doi [Universe] {10.3390/universe7030056}, \href
  {https://ui.adsabs.harvard.edu/abs/2021Univ....7...56L} {7, 56}

\bibitem[\protect\citeauthoryear{{Macquart} et~al.,}{{Macquart}
  et~al.}{2010}]{Macquart:2010}
{Macquart} J.-P.,  et~al., 2010, \mn@doi [\pasa] {10.1071/AS09082}, \href
  {https://ui.adsabs.harvard.edu/abs/2010PASA...27..272M} {27, 272}

\bibitem[\protect\citeauthoryear{{Majid}, {Pearlman}, {Nimmo}, {Hessels},
  {Prince}, {Naudet}, {Kocz}  \& {Horiuchi}}{{Majid} et~al.}{2020}]{Majid:2020}
{Majid} W.~A.,  {Pearlman} A.~B.,  {Nimmo} K.,  {Hessels} J. W.~T.,  {Prince}
  T.~A.,  {Naudet} C.~J.,  {Kocz} J.,   {Horiuchi} S.,  2020, \mn@doi [\apjl]
  {10.3847/2041-8213/ab9a4a}, \href
  {https://ui.adsabs.harvard.edu/abs/2020arXiv200406845M} {897, L4}

\bibitem[\protect\citeauthoryear{{Marcote} et~al.,}{{Marcote}
  et~al.}{2020}]{Marcote:2020}
{Marcote} B.,  et~al., 2020, \mn@doi [\nat] {10.1038/s41586-019-1866-z}, \href
  {https://ui.adsabs.harvard.edu/abs/2020Natur.577..190M} {577, 190}

\bibitem[\protect\citeauthoryear{{Marcote} et~al.,}{{Marcote}
  et~al.}{2021}]{Marcote:2021ATel14603}
{Marcote} B.,  et~al., 2021, The Astronomer's Telegram, \href
  {https://ui.adsabs.harvard.edu/abs/2021ATel14603....1M} {14603, 1}

\bibitem[\protect\citeauthoryear{{Marthi}, {Gautam}, {Li}, {Lin}, {Main},
  {Naidu}, {Pen}  \& {Wharton}}{{Marthi} et~al.}{2020}]{Marthi:2020}
{Marthi} V.~R.,  {Gautam} T.,  {Li} D.~Z.,  {Lin} H.~H.,  {Main} R.~A.,
  {Naidu} A.,  {Pen} U.~L.,   {Wharton} R.~S.,  2020, \mn@doi [\mnras]
  {10.1093/mnrasl/slaa148}, \href
  {https://ui.adsabs.harvard.edu/abs/2020MNRAS.499L..16M} {499, L16}

\bibitem[\protect\citeauthoryear{{Marthi} et~al.,}{{Marthi}
  et~al.}{2022}]{Marthi:2021}
{Marthi} V.~R.,  et~al., 2022, \mn@doi [\mnras] {10.1093/mnras/stab3067}, \href
  {https://ui.adsabs.harvard.edu/abs/2021MNRAS.tmp.2782M} {509, 2209}

\bibitem[\protect\citeauthoryear{{Masui} et~al.,}{{Masui}
  et~al.}{2015}]{Masui:2015}
{Masui} K.,  et~al., 2015, \mn@doi [\nat] {10.1038/nature15769}, \href
  {https://ui.adsabs.harvard.edu/abs/2015Natur.528..523M} {528, 523}

\bibitem[\protect\citeauthoryear{{Metzger}, {Berger}  \& {Margalit}}{{Metzger}
  et~al.}{2017}]{Metzger:2017}
{Metzger} B.~D.,  {Berger} E.,   {Margalit} B.,  2017, \mn@doi [\apj]
  {10.3847/1538-4357/aa633d}, \href
  {https://ui.adsabs.harvard.edu/abs/2017ApJ...841...14M} {841, 14}

\bibitem[\protect\citeauthoryear{{Metzger}, {Margalit}  \& {Sironi}}{{Metzger}
  et~al.}{2019}]{Metzger:2019}
{Metzger} B.~D.,  {Margalit} B.,   {Sironi} L.,  2019, \mn@doi [\mnras]
  {10.1093/mnras/stz700}, \href
  {https://ui.adsabs.harvard.edu/abs/2019MNRAS.485.4091M} {485, 4091}

\bibitem[\protect\citeauthoryear{{Michilli} et~al.,}{{Michilli}
  et~al.}{2018}]{Michilli:2018}
{Michilli} D.,  et~al., 2018, \mn@doi [\nat] {10.1038/nature25149}, \href
  {https://ui.adsabs.harvard.edu/abs/2018Natur.553..182M} {553, 182}

\bibitem[\protect\citeauthoryear{{Newville}, {Stensitzki}, {Allen}, {Rawlik},
  {Ingargiola}  \& {Nelson}}{{Newville} et~al.}{2016}]{Newville:2016}
{Newville} M.,  {Stensitzki} T.,  {Allen} D.~B.,  {Rawlik} M.,  {Ingargiola}
  A.,   {Nelson} A.,  2016, {Lmfit: Non-Linear Least-Square Minimization and
  Curve-Fitting for Python} (\mn@eprint {ascl} {1606.014})

\bibitem[\protect\citeauthoryear{{Nimmo} et~al.,}{{Nimmo}
  et~al.}{2021}]{Nimmo:2021}
{Nimmo} K.,  et~al., 2021, \mn@doi [Nature Astronomy]
  {10.1038/s41550-021-01321-3}, \href
  {https://ui.adsabs.harvard.edu/abs/2021NatAs...5..594N} {5, 594}

\bibitem[\protect\citeauthoryear{{Noutsos}, {Johnston}, {Kramer}  \&
  {Karastergiou}}{{Noutsos} et~al.}{2008}]{Noutsos:2008}
{Noutsos} A.,  {Johnston} S.,  {Kramer} M.,   {Karastergiou} A.,  2008, \mn@doi
  [\mnras] {10.1111/j.1365-2966.2008.13188.x}, \href
  {https://ui-adsabs-harvard-edu.ezproxy.lib.swin.edu.au/abs/2008MNRAS.386.1881N}
  {386, 1881}

\bibitem[\protect\citeauthoryear{{Noutsos}, {Karastergiou}, {Kramer},
  {Johnston}  \& {Stappers}}{{Noutsos} et~al.}{2009}]{Noutsos:2009}
{Noutsos} A.,  {Karastergiou} A.,  {Kramer} M.,  {Johnston} S.,   {Stappers}
  B.~W.,  2009, \mn@doi [\mnras] {10.1111/j.1365-2966.2009.14806.x}, \href
  {https://ui.adsabs.harvard.edu/abs/2009MNRAS.396.1559N} {396, 1559}

\bibitem[\protect\citeauthoryear{{O'Connor}, {Piro}, {Lotti}, {Ricci}, {Bruni}
  \& {Zhang}}{{O'Connor} et~al.}{2021}]{O'Connor:2021ATel14525}
{O'Connor} B.,  {Piro} L.,  {Lotti} S.,  {Ricci} R.,  {Bruni} G.,   {Zhang} B.,
   2021, The Astronomer's Telegram, \href
  {https://ui.adsabs.harvard.edu/abs/2021ATel14525....1O} {14525, 1}

\bibitem[\protect\citeauthoryear{{Oppermann} et~al.,}{{Oppermann}
  et~al.}{2015}]{Oppermann:2015}
{Oppermann} N.,  et~al., 2015, \mn@doi [\aap] {10.1051/0004-6361/201423995},
  \href {https://ui.adsabs.harvard.edu/abs/2015A&A...575A.118O} {575, A118}

\bibitem[\protect\citeauthoryear{{Os{\l}owski} et~al.,}{{Os{\l}owski}
  et~al.}{2019}]{Oslowski:2019}
{Os{\l}owski} S.,  et~al., 2019, \mn@doi [\mnras] {10.1093/mnras/stz1751},
  \href {https://ui.adsabs.harvard.edu/abs/2019MNRAS.488..868O} {488, 868}

\bibitem[\protect\citeauthoryear{{Pastor-Marazuela} et~al.,}{{Pastor-Marazuela}
  et~al.}{2021}]{Marazuela:2021}
{Pastor-Marazuela} I.,  et~al., 2021, \mn@doi [\nat]
  {10.1038/s41586-021-03724-8}, \href
  {https://ui.adsabs.harvard.edu/abs/2021Natur.596..505P} {596, 505}

\bibitem[\protect\citeauthoryear{{Pearlman}, {Majid}, {Prince}, {Bansal},
  {Naudet}  \& {Kocz}}{{Pearlman} et~al.}{2021}]{Pearlman:2021ATel14519}
{Pearlman} A.~B.,  {Majid} W.~A.,  {Prince} T.~A.,  {Bansal} K.,  {Naudet}
  C.~J.,   {Kocz} J.,  2021, The Astronomer's Telegram, \href
  {https://ui.adsabs.harvard.edu/abs/2021ATel14519....1P} {14519, 1}

\bibitem[\protect\citeauthoryear{{Petroff} et~al.,}{{Petroff}
  et~al.}{2015}]{Petroff:2015}
{Petroff} E.,  et~al., 2015, \mn@doi [\mnras] {10.1093/mnras/stu2419}, \href
  {https://ui.adsabs.harvard.edu/abs/2015MNRAS.447..246P} {447, 246}

\bibitem[\protect\citeauthoryear{{Petroff}, {Hessels}  \& {Lorimer}}{{Petroff}
  et~al.}{2022}]{Petroff:2022}
{Petroff} E.,  {Hessels} J.~W.~T.,   {Lorimer} D.~R.,  2022, \mn@doi [\aapr]
  {10.1007/s00159-022-00139-w}, \href
  {https://ui.adsabs.harvard.edu/abs/2021arXiv210710113P} {30, 2}

\bibitem[\protect\citeauthoryear{{Piro} et~al.,}{{Piro}
  et~al.}{2021}]{Piro:2021}
{Piro} L.,  et~al., 2021, \mn@doi [\aap] {10.1051/0004-6361/202141903}, \href
  {https://ui.adsabs.harvard.edu/abs/2021A&A...656L..15P} {656, L15}

\bibitem[\protect\citeauthoryear{{Platts}, {Weltman}, {Walters}, {Tendulkar},
  {Gordin}  \& {Kandhai}}{{Platts} et~al.}{2019}]{Platts:2019}
{Platts} E.,  {Weltman} A.,  {Walters} A.,  {Tendulkar} S.~P.,  {Gordin}
  J.~E.~B.,   {Kandhai} S.,  2019, \mn@doi [\physrep]
  {10.1016/j.physrep.2019.06.003}, \href
  {https://ui.adsabs.harvard.edu/abs/2019PhR...821....1P} {821, 1}

\bibitem[\protect\citeauthoryear{{Platts} et~al.,}{{Platts}
  et~al.}{2021}]{Platts:2021}
{Platts} E.,  et~al., 2021, \mn@doi [\mnras] {10.1093/mnras/stab1544}, \href
  {https://ui.adsabs.harvard.edu/abs/2021MNRAS.505.3041P} {505, 3041}

\bibitem[\protect\citeauthoryear{{Pleunis} et~al.,}{{Pleunis}
  et~al.}{2021a}]{Pleunis:2021_LOFAR}
{Pleunis} Z.,  et~al., 2021a, \mn@doi [\apjl] {10.3847/2041-8213/abec72}, \href
  {https://ui.adsabs.harvard.edu/abs/2021ApJ...911L...3P} {911, L3}

\bibitem[\protect\citeauthoryear{{Pleunis} et~al.,}{{Pleunis}
  et~al.}{2021b}]{Pleunis:2021}
{Pleunis} Z.,  et~al., 2021b, \mn@doi [\apj] {10.3847/1538-4357/ac33ac}, \href
  {https://ui.adsabs.harvard.edu/abs/2021ApJ...923....1P} {923, 1}

\bibitem[\protect\citeauthoryear{{Radhakrishnan} \& {Rankin}}{{Radhakrishnan}
  \& {Rankin}}{1990}]{Radhakrishnan:1990}
{Radhakrishnan} V.,  {Rankin} J.~M.,  1990, \mn@doi [\apj] {10.1086/168531},
  \href {https://ui.adsabs.harvard.edu/abs/1990ApJ...352..258R} {352, 258}

\bibitem[\protect\citeauthoryear{{Rajwade} et~al.,}{{Rajwade}
  et~al.}{2020}]{Rajwade:2020}
{Rajwade} K.~M.,  et~al., 2020, \mn@doi [\mnras] {10.1093/mnras/staa1237},
  \href {https://ui.adsabs.harvard.edu/abs/2020MNRAS.495.3551R} {495, 3551}

\bibitem[\protect\citeauthoryear{{Ravi} et~al.,}{{Ravi}
  et~al.}{2019}]{Ravi:2019_localization}
{Ravi} V.,  et~al., 2019, \mn@doi [\nat] {10.1038/s41586-019-1389-7}, \href
  {https://ui.adsabs.harvard.edu/abs/2019Natur.572..352R} {572, 352}

\bibitem[\protect\citeauthoryear{{Ravi} et~al.,}{{Ravi}
  et~al.}{2021}]{Ravi:2021}
{Ravi} V.,  et~al., 2021, arXiv e-prints, \href
  {https://ui.adsabs.harvard.edu/abs/2021arXiv210609710R} {p. arXiv:2106.09710}

\bibitem[\protect\citeauthoryear{{Ricci}, {Piro}, {Panessa}, {O'Connor},
  {Lotti}, {Bruni}  \& {Zhang}}{{Ricci} et~al.}{2021}]{Ricci:2021ATel14549}
{Ricci} R.,  {Piro} L.,  {Panessa} F.,  {O'Connor} B.,  {Lotti} S.,  {Bruni}
  G.,   {Zhang} B.,  2021, The Astronomer's Telegram, \href
  {https://ui.adsabs.harvard.edu/abs/2021ATel14549....1R} {14549, 1}

\bibitem[\protect\citeauthoryear{{Safarzadeh}, {Prochaska}, {Heintz}  \&
  {Fong}}{{Safarzadeh} et~al.}{2020}]{Safarzadeh:2020}
{Safarzadeh} M.,  {Prochaska} J.~X.,  {Heintz} K.~E.,   {Fong} W.-f.,  2020,
  \mn@doi [\apjl] {10.3847/2041-8213/abd03e}, \href
  {https://ui.adsabs.harvard.edu/abs/2020ApJ...905L..30S} {905, L30}

\bibitem[\protect\citeauthoryear{{Seymour}, {Michilli}  \& {Pleunis}}{{Seymour}
  et~al.}{2019}]{Seymour:2019}
{Seymour} A.,  {Michilli} D.,   {Pleunis} Z.,  2019, {DM\_phase: Algorithm for
  correcting dispersion of radio signals} (\mn@eprint {ascl} {1910.004})

\bibitem[\protect\citeauthoryear{{Shannon} et~al.,}{{Shannon}
  et~al.}{2018}]{Shannon:2018}
{Shannon} R.~M.,  et~al., 2018, \mn@doi [\nat] {10.1038/s41586-018-0588-y},
  \href {https://ui.adsabs.harvard.edu/abs/2018Natur.562..386S} {562, 386}

\bibitem[\protect\citeauthoryear{{Spitler} \& {Hilmarsson}}{{Spitler} \&
  {Hilmarsson}}{2021}]{Spitler:2021ATel14537}
{Spitler} L.,  {Hilmarsson} H.,  2021, The Astronomer's Telegram, \href
  {https://ui.adsabs.harvard.edu/abs/2021ATel14537....1S} {14537, 1}

\bibitem[\protect\citeauthoryear{{Spitler} et~al.,}{{Spitler}
  et~al.}{2016}]{Spitler:2016}
{Spitler} L.~G.,  et~al., 2016, \mn@doi [\nat] {10.1038/nature17168}, \href
  {https://ui.adsabs.harvard.edu/abs/2016Natur.531..202S} {531, 202}

\bibitem[\protect\citeauthoryear{{Sridhar}, {Metzger}, {Beniamini}, {Margalit},
  {Renzo}, {Sironi}  \& {Kovlakas}}{{Sridhar} et~al.}{2021}]{Sridhar:2021}
{Sridhar} N.,  {Metzger} B.~D.,  {Beniamini} P.,  {Margalit} B.,  {Renzo} M.,
  {Sironi} L.,   {Kovlakas} K.,  2021, \mn@doi [\apj]
  {10.3847/1538-4357/ac0140}, \href
  {https://ui.adsabs.harvard.edu/abs/2021ApJ...917...13S} {917, 13}

\bibitem[\protect\citeauthoryear{{Tendulkar} et~al.,}{{Tendulkar}
  et~al.}{2021}]{Tendulkar:2021}
{Tendulkar} S.~P.,  et~al., 2021, \mn@doi [\apjl] {10.3847/2041-8213/abdb38},
  \href {https://ui.adsabs.harvard.edu/abs/2021ApJ...908L..12T} {908, L12}

\bibitem[\protect\citeauthoryear{{Wharton} et~al.,}{{Wharton}
  et~al.}{2021a}]{Wharton:2021ATel14529}
{Wharton} R.,  et~al., 2021a, The Astronomer's Telegram, \href
  {https://ui.adsabs.harvard.edu/abs/2021ATel14529....1W} {14529, 1}

\bibitem[\protect\citeauthoryear{{Wharton} et~al.,}{{Wharton}
  et~al.}{2021b}]{Wharton:2021ATel14538}
{Wharton} R.,  et~al., 2021b, The Astronomer's Telegram, \href
  {https://ui.adsabs.harvard.edu/abs/2021ATel14538....1W} {14538, 1}

\bibitem[\protect\citeauthoryear{{Xu} et~al.,}{{Xu}
  et~al.}{2021}]{Xu:2021ATel14518}
{Xu} H.,  et~al., 2021, The Astronomer's Telegram, \href
  {https://ui.adsabs.harvard.edu/abs/2021ATel14518....1X} {14518, 1}

\bibitem[\protect\citeauthoryear{{Yang}, {Li}  \& {Zhang}}{{Yang}
  et~al.}{2020}]{Yang:2020}
{Yang} Y.-P.,  {Li} Q.-C.,   {Zhang} B.,  2020, \mn@doi [\apj]
  {10.3847/1538-4357/ab88ab}, \href
  {https://ui.adsabs.harvard.edu/abs/2020ApJ...895....7Y} {895, 7}

\bibitem[\protect\citeauthoryear{{Zhang}, {Wang}, {Wu}, {Wang}, {Li}, {Dai}  \&
  {Zhang}}{{Zhang} et~al.}{2021}]{Zhang:2021}
{Zhang} G.~Q.,  {Wang} P.,  {Wu} Q.,  {Wang} F.~Y.,  {Li} D.,  {Dai} Z.~G.,
  {Zhang} B.,  2021, \mn@doi [\apjl] {10.3847/2041-8213/ac2a3b}, \href
  {https://ui.adsabs.harvard.edu/abs/2021ApJ...920L..23Z} {920, L23}

\bibitem[\protect\citeauthoryear{{Zhirkov} et~al.,}{{Zhirkov}
  et~al.}{2021}]{Zhirkov:2021ATel14532}
{Zhirkov} K.,  et~al., 2021, The Astronomer's Telegram, \href
  {https://ui.adsabs.harvard.edu/abs/2021ATel14532....1Z} {14532, 1}

\bibitem[\protect\citeauthoryear{{van Straten}}{{van
  Straten}}{2004}]{vanStraten:2004}
{van Straten} W.,  2004, \mn@doi [\apjs] {10.1086/383187}, \href
  {https://ui.adsabs.harvard.edu/abs/2004ApJS..152..129V} {152, 129}

\bibitem[\protect\citeauthoryear{{van Straten} \& {Bailes}}{{van Straten} \&
  {Bailes}}{2011}]{dspsr}
{van Straten} W.,  {Bailes} M.,  2011, \mn@doi [\pasa] {10.1071/AS10021}, \href
  {https://ui.adsabs.harvard.edu/abs/2011PASA...28....1V} {28, 1}

\bibitem[\protect\citeauthoryear{{van der Velden}}{{van der
  Velden}}{2020}]{Velden:2020_cmasher}
{van der Velden} E.,  2020, \mn@doi [The Journal of Open Source Software]
  {10.21105/joss.02004}, \href
  {https://ui.adsabs.harvard.edu/abs/2020JOSS....5.2004V} {5, 2004}

\makeatother
\end{thebibliography}




\bsp	
\label{lastpage}
\end{document}